\documentclass[aps, prb, twocolumn, nofootinbib, superscriptaddress, longbibliography,  floatfix]{revtex4-2}

\pdfoutput=1
\usepackage[utf8]{inputenc}
\DeclareUnicodeCharacter{00A0}{ }
\usepackage{amsmath}
\usepackage{amssymb}
\usepackage{physics}
\usepackage[makeroom]{cancel}
\usepackage{stmaryrd}

\usepackage{graphicx}
\usepackage{color}
\usepackage[english]{babel}
\usepackage{stmaryrd}
\usepackage{microtype}

\usepackage[percent]{overpic}
\usepackage{tikz}
\usetikzlibrary{calc,shadows,matrix,positioning,patterns,decorations.pathreplacing}

\usepackage{hyperref}
\hypersetup{
	pdftoolbar=true,
	pdfmenubar=true,
	pdffitwindow=false,
	pdfstartview={FitH},
	pdftitle={},
	pdfauthor={Bertrand, Bauernfeind, Dumitrescu, Maček, Parcollet, Waintal},
	pdfsubject={},
	pdfcreator={},
	pdfproducer={},
	pdfnewwindow=true,
	colorlinks=true,
	linkcolor=black,
	citecolor=blue,
	filecolor=magenta,
	urlcolor=blue
}

\usepackage{xr}

\renewcommand{\vec}{\boldsymbol}

\newcommand{\phdag}{{\phantom{\dagger}}}

\newcommand*{\lists}[2]{\left\llbracket \begin{matrix} #1 \\ #2 \end{matrix} \right\rrbracket}

\newcommand*{\expHdt}[1]{e^{-i #1 H \Delta t}}
\newcommand*{\expHdthalf}[1]{e^{-i #1 H \Delta t/2}}

\usepackage{xcolor}

\newcommand{\eqnref}[1]{Eq.~\eqref{#1}}

\newcommand{\CCQ}{Center for Computational Quantum Physics, Flatiron Institute, 162 5th Avenue, New York, NY 10010, USA}
\newcommand{\Grenoble}{Universit\'e Grenoble  Alpes,  CEA,  IRIG-PHELIQS,  38000  Grenoble, France}

\begin{document}
\title{Quantum Quasi-Monte Carlo algorithm for out-of-equilibrium Green functions at
   long times}

\author{Corentin Bertrand}
\email{cbertrand@flatironinstitute.org}
\affiliation{\CCQ}

\author{Daniel Bauernfeind}
\affiliation{\CCQ}

\author{Philipp T.~Dumitrescu}
\affiliation{\CCQ}

\author{Marjan Maček}
\affiliation{\Grenoble}

\author{Xavier Waintal}
\affiliation{\Grenoble}

\author{Olivier Parcollet}
\affiliation{\CCQ}
\affiliation{Universit\'e Paris-Saclay, CNRS, CEA, Institut de physique th\'eorique, 91191,
   Gif-sur-Yvette, France}

\date{\today}

\begin{abstract}
   We extend the recently developed Quantum Quasi-Monte Carlo (QQMC) approach to
   obtain the full frequency dependence of Green functions in a single
   calculation. QQMC is a general approach for calculating high-order perturbative
   expansions in power of the electron-electron interaction strength. In contrast
   to conventional Markov chain Monte Carlo sampling, QQMC uses low-discrepancy
   sequences for a more uniform sampling of the multi-dimensional integrals
   involved and can potentially outperform Monte Carlo by several orders of
   magnitudes. A core concept of QQMC is the a priori construction of a ``model
   function" that approximates the integrand and is used to optimize the sampling
   distribution. In this paper, we show that the model function concept extends to
   a kernel approach for the computation of Green functions. We illustrate the
   approach on the Anderson impurity model and show that the scaling of the error
   with the number of integrand evaluations $N$ is
   $\sim
   1/N^{0.86}$ in the best cases, and comparable to Monte Carlo scaling
   $\sim
   1/N^{0.5}$ in the worst cases. We find a systematic improvement over
   Monte Carlo sampling by at least two orders of magnitude while using a basic
   form of model function. Finally, we compare QQMC results with calculations
   performed with the Fork Tensor Product State (FTPS) method, a recently
   developed tensor network approach for solving impurity problems. Applying a
   simple Padé approximant for the series resummation, we find that QQMC matches
   the FTPS results beyond the perturbative regime.
\end{abstract}

\maketitle

\section{Introduction}

Despite considerable advances in numerical approaches to condensed matter
systems, many algorithms still lack the control or precision necessary to study
strongly correlated phenomena. At the same time, recent experimental
developments have allowed unprecedented precision in characterizing quantum
many-body states -- in systems as varied as atomic
gases~\cite{Gross_Bloch_2017}, trapped Rydberg atoms~\cite{Lukin_51_atoms_2017}
trapped ions~\cite{Blatt_Ross_TrappedIons_2012}, nano-electronic
devices~\cite{Goldhaber-Gordon1998, Goldhaber-Gordon1998a,
Kouwenhoven_Kondo_2000, Pierre_Science_Kondo_2018} --  where quantitative numerical predictions
can provide valuable comparisons and give insights into new physics. To
overcome limitations in precision, studying interacting quantum many-body
systems by numerically evaluating high-order perturbation series and applying
resummation techniques has seen recently seen unexpected renewed interest
\cite{Prokofev_9804, Prokofev_0801, Mishchenko_9910,
VanHoucke_1110, Profumo_1504, Wu_1608, Rossi_1612, Chen_1809,
Bertrand_1903_series, Bertrand_1903_kernel, Moutenet_1904,
Rossi_2001, Macek2020}.

Among the various regimes of strongly correlated systems, calculating dynamical
properties at long times or low frequencies has been particularly challenging
for numerical approaches. This applies both to calculating real-frequency
correlation functions of equilibrium systems and to out-of-equilibrium systems
-- such as ones subjected to strong driving fields or external currents.
Imaginary time algorithms require ill-conditioned analytical continuations to
extract real-time properties. Real-time algorithms face intrinsic limitations
to reaching long-time behavior, such as the dynamical sign problem for Monte
Carlo methods or prohibitive entanglement growth for Tensor Network
methods~\cite{Calabrese_2005}.
To address the challenging long-time regime, we have recently developed a new
approach~\cite{Profumo_1504,Bertrand_1903_series,Bertrand_1903_kernel,Macek2020} based on high order real-time
Schwinger-Keldysh perturbation theory.

In Ref.~\onlinecite{Macek2020}, we most recently introduced the ``Quantum
Quasi-Monte Carlo'' (QQMC) method. This is based on calculating the integrals
in perturbation series coefficients by using low-discrepancy sequences rather
than conventional Monte Carlo sampling. We demonstrated a dramatic increase in
performance due to improved algorithmic scaling of this method, with
convergence as fast as $\sim 1/N$ in the number of samples
$N$. We applied QQMC to compute observables for the Anderson
impurity model both in and out of equilibrium, and were able to quickly sweep a
large range of parameters. While the approach of Ref.~\onlinecite{Macek2020}
is general, it relies on the concept of a ``model function'', which serves the
role analogous to importance sampling in traditional Monte Carlo methods and
incorporates a priori knowledge of the integrand. Unlike Monte Carlo sampling,
however, in QQMC, the rate of convergence with $N$  itself
depends on the choice of model function and can be improved with additional a
priori knowledge. This raises the question: how well can QQMC be applied to
more complex observables than the previously studied local densities and
currents?

In this paper, we address this question by adapting QQMC to the problem of
calculating the full frequency-dependent Green function of the Anderson
impurity  model. We use the kernel approach of Ref.~\onlinecite{Bertrand_1903_kernel}, which computes the frequency dependence by
integrating with a single sampling for each perturbative coefficient. We
develop an automated way to obtain simple effective model functions for these
integrals, in the form of a product of one dimensional functions. We compare
the efficiency of applying QQMC in a kernel approach to performing separate
calculations at individual frequencies, and remarkably find no advantage in
separating frequencies. In spite of its simplicity, a single model function
with a single sequence of points can compute a continuum of integrals with
convergence rates that are systematically better than $1/\sqrt{N}$
(standard Monte Carlo sampling) and as high as $\sim 1/N^{0.86}$. In
practice, applying QQMC to compute coefficients up to order 10, is 2-3 orders
of magnitude faster than the Monte Carlo approach of
Ref.~\onlinecite{Bertrand_1903_kernel}.

From the perturbation series coefficients, we compute the Green function at
large interactions strengths, by using Padé approximants for series
resummation~\cite{baker1996pade}. We compare the QQMC result with
calculations using the Fork Tensor Product State (FTPS) solver
\cite{BauernfeindFTPSorig}. This solver uses a particular ``fork" Tensor Network
(TN) to represent quantum states of impurity models and performs the real-time
evolution of such states. It provides a non-perturbative way to obtain
real-time Green functions which is fundamentally different from QQMC. We find
excellent agreement between the two methods. Throughout, we will discuss
technical developments and show how algorithmic choices affect the
computational performance.

This article is organized as follows. Section~\ref{sec:model} presents
the Anderson impurity model used in this article. Sec.~\ref{sec:QQMC-full}
focuses on the QQMC algorithm. We introduce the kernel formalism
(Sec.~\ref{sec:definitions}), describe the QQMC method and provide two
algorithms which adapts QQMC into computing Green functions
(Sec.~\ref{sec:QQMC}). After explaining how to obtain a model function
(Sec.~\ref{sec:projection_warper}), we discuss the performance and convergence
rates of the new QQMC methods (Sec.~\ref{sec:convergence}). Next, we focus on
comparing our results to FTPS. We detail the resummation by Padé approximant in
Sec.\ref{sec:pade}. Technical aspects of FTPS are given in
Sec.~\ref{sec:FTPS}. Finally, the results of the comparison are
discussed in Sec.~\ref{sec:comparison}.

\section{Model}
\label{sec:model}

Although QQMC is applicable to a general -- potentially non-equilibrium --
system, we will focus the discussion on the equilibrium single band Anderson
impurity model \cite{Anderson1961} for simplicity. We consider an impurity
experiencing on-site Coulomb repulsion, symmetrically coupled to two identical
leads with semi-circular density of states. It can be represented by a
one-dimensional infinite chain of electronic sites with Hamiltonian
\begin{equation}\label{eq:Htot}
   H(t) = H_{0} +
   H_{\mathrm{int}}\theta(t),
\end{equation}
where
\begin{align}
   \label{eq:H0}
   H_{0}            & = \sum_{x,\sigma} \left( \gamma_x c^{\dag}_{x,\sigma} c^{\phdag}_{x+1,\sigma}
   + \mathrm{H.c.} \right)+ E_d  \sum_\sigma c^{\dag}_{0\sigma} c^{\phdag}_{0\sigma},               \\
   \label{eq:Hint}
   H_{\mathrm{int}} & = U (c^{\dag}_{0\uparrow} c^{\phdag}_{0\uparrow} - \alpha)
   (c^{\dag}_{0\downarrow} c^{\phdag}_{0\downarrow} - \alpha),
\end{align}
and $\theta(t)$ is the Heaviside step function.
This Hamiltonian describes an interacting impurity at site
$x=0$ coupled to two non-interacting leads, corresponding to
sites $x < 0$ and $x > 0$. Here
$\sigma = \uparrow, \downarrow$ denotes the electronic spin. The electron hopping term
between the impurity and the last site of each lead is $\gamma_0 = \gamma_{-1} = \gamma$.
Within each lead, the hopping term between sites is constant
$\gamma_x = D/2$, so that the leads have a semi-circular density of
states with half-bandwidth $D$.

As is standard, the effects of the leads on the impurity are encoded in a
hybridization function $\Delta(\omega)$. For \eqnref{eq:H0}, the
retarded non-interacting Green function of the impurity is
$g^R(\omega) = 1 / (\omega - E_d - \Delta^R(\omega)) $, with
\begin{equation}
   \Delta^{R}(\omega) = \frac{\Gamma}{D} \cdot
   \begin{cases}
      \left( \omega + \sqrt{\omega^2 - D^2} \right),   & \omega < -D,           \\
      \left( \omega - i \sqrt{D^2 - \omega^2} \right), & -D \leq \omega \leq D, \\
      \left( \omega - \sqrt{\omega^2 - D^2} \right),   & \omega > D.
   \end{cases}
\end{equation}
Here we have defined the tunneling rate from the impurity to the leads at the
equilibrium Fermi level $\Gamma = 4\gamma^2 / D$. The leads are half filled and
at zero temperature. We use units such that $\hbar = 1$.

In the system described by \eqnref{eq:Htot}, the local Coulomb repulsion
on the impurity $U$ is quenched on at
$t=0$.  We have introduced a quadratic shift in
$H_{\mathrm{int}}$ parameterized by $\alpha$. This shift is
compensated by the $E_d$ term in $H_0$, so
that the energy of the impurity charged with a single electron is
$E_d - \alpha U$ after the quench. Performing calculations at non-zero
$\alpha$ changes the expansion point of the perturbation series
and can be useful in improving series convergence~\cite{Profumo_1504,Rubtsov2004,Wu_1608}. In
this work we use $\Gamma = 1/2$, $D=11.476\Gamma$ (the value was
chosen to facilitate bath discretization in FTPS) and $\alpha = 1/2$.
We will consider two models: $E_d = 0$ which is particle--hole
symmetric at all $U$, and $E_d = \Gamma$ which
breaks this symmetry.

\section{Green Function Calculation with QQMC}
\label{sec:QQMC-full}

\subsection{Summary of Diagrammatic Expansions and the Kernel Approach}
\label{sec:definitions}

Our algorithms are based on real-time perturbation theory and the kernel
approach of Ref.~\onlinecite{Bertrand_1903_kernel}. Here, we briefly recall the relevant
previous result, but refer to Ref.~\onlinecite{Bertrand_1903_kernel} for a more
comprehensive description. While the kernel approach was formulated for general
models and interactions, in this article we directly specialize our discussion
to the Anderson impurity model \eqnref{eq:Htot}.

Our method is based on performing a perturbative expansion of the real-time
Green function $G^{ab}(t,t')$. Here $a,b\in\{0,1\}$ are the
Keldysh contour indices, so that
\begin{equation}
   G^{ab}(t,t') =
   \begin{pmatrix}
      G^T(t,t') & G^<(t,t')               \\[0.5em]
      G^>(t,t') & G^{\widetilde{T}}(t,t')
   \end{pmatrix}_{ab},
   \label{eq:gf_def}
\end{equation}
where $G^T(t,t')$, $G^<(t,t')$, $G^>(t,t')$
and $G^{\widetilde{T}}(t,t')$ are respectively the time ordered, lesser, greater
and anti-time ordered impurity Green functions. We denote the non-interacting
impurity Green function $g^{ab}(t,t')$. For notational simplicity, we
also define combined indices $X = (t, a)$, $Y=(t', b)$
to write expressions such as $g(X, Y) = g^{ab}(t,t')$ or
$\delta(X,Y) =
\delta(t-t')\delta_{ab}$. The interacting system \eqnref{eq:Htot} is spin
symmetric and we suppress the Green function spin indices
$\sigma,\sigma'$ throughout, noting that $G_{\uparrow\uparrow}(X,Y) = G_{\downarrow\downarrow}(X,Y)$ and
$G_{\uparrow\downarrow}(X,Y) = G_{\downarrow\uparrow}(X,Y) = 0$. Additionally, we will only consider the impurity Green
function itself, although the approach is straightforward to generalize to
multiple electron sites or orbitals.

The Schwinger-Keldysh perturbation series for the Green function in powers of
$U$ is~\cite{Rammer_2007}
\begin{multline}
   \label{eq:integral_gf}
   G^{ab}(t,t') = \sum_{n = 0}^{\infty}
   \frac{i^nU^n}{n!} \int_0^{t_M} \!\!\dd{u_1} \ldots
   \dd{u_n} \Biggl\{
   \sum_{\{a_k\}} (-1)^{\sum_k a_k}	\\
   \cdot \lists{( t, a), U_1, \ldots, U_n}{(t', b), U_1, \ldots, U_n}
   \lists{U_1, \ldots, U_n}{U_1, \ldots, U_n} \Biggr\}.
\end{multline}
Here $U_k = (u_k, a_k)$ are the coordinates of the interaction vertices
located on the impurity, at time $u_k$ and with Keldysh index
$a_k \in \{0, 1\}$. The times $u_k$ are integrated from
$t=0$, when interaction is quenched on, to the time of
measurement $t_M \ge t, t'$. We have also adopted the notation of
Ref.~\onlinecite{Bertrand_1903_kernel} for Wick determinants
\begin{equation}
   \label{eq:wick_det}
   \lists{A_1, \ldots, A_m}{B_1, \ldots, B_m} =
   \begin{vmatrix}
      {g}(A_1, B_1) & \hdots & {g}(A_1, B_m) \\
      \vdots        & \ddots & \vdots        \\
      {g}(A_m, B_1) & \hdots & {g}(A_m, B_m)
   \end{vmatrix},
\end{equation}
where $A_i$ and $B_j$ are combined indices
of a time and a Keldysh index. In \eqnref{eq:integral_gf}, the first Wick
determinant corresponds to one species of spin, while the second corresponds to
the other.

There are two technical aspects in the perturbative expansion that are
suppressed in the notation of \eqnref{eq:wick_det}. First, the Green
functions $g(U_k, U_k)$ which are on the diagonal of the determinant
and correspond to interaction vertices $U_k$ are replaced by
$g^<(u_k, u_k) - i \alpha$. The choice of $g^<$ reflects the
operator ordering in the interaction Hamiltonian \eqnref{eq:Hint}, while
the $\alpha$ term reflects the quadratic shift
\cite{Profumo_1504}. Second, Green functions $g^{T}$ and
$g^{\widetilde{T}}$ have a discontinuity at equal times. Their value here
should respect the convention taken when defining the time-ordering operator
(see Appendix~\ref{app:time-splitting}).

A direct evaluation of \eqnref{eq:integral_gf} would give the Green function
only at a single pair of fixed times $t, t'$. In order to
compute the entire time dependence at once, Ref.~\onlinecite{Bertrand_1903_kernel}
defined a kernel $K^{cb}(u,t') = K(Z, Y)$, with $Z=(u,c)$, such
that
\begin{equation}\label{eq:G_Keldysh_expansion3}
   G(X,Y) = g(X,Y) + \sum_{Z} (-1)^c g(X,Z) K(Z,Y),
\end{equation}
where $\sum_{Z} = \int \dd{u} \sum_{c}$. The explicit expression for
$K(Z,Y)$ is found by expanding the first determinant in
\eqnref{eq:integral_gf} by minors along the first row
\begin{widetext}
   \begin{align}\label{eq:kernel}
      K(Z,Y)                           & =
      \sum^\infty_{n = 1} U^n
      \int \dd{u_1} \ldots \dd{u_n} f_n(Z, Y, \vec{u}),
      \\
      f_n(Z,Y,\vec{u})                 & = (-1)^c \sum_{p=1}^{n}
      \sum_{a_p}
      (-1)^{a_p} \delta\left(Z, U_p\right)
      W^n_p\left(Y,\vec{u}, a_p\right),
      \\
      W^n_p\left(Y,\vec{u}, a_p\right) & =
      \frac{i^n}{n!}
      \sum_{ \{a_k\} \atop k\neq p}
      (-1)^{\sum_{k\neq p}a_k} \cdot
      (-1)^p
      \lists{\;\;U_1\;\;,\, \dotfill, U_n}{(t', b), \ldots, \cancel{U_p}, \ldots, U_n}
      \lists{U_1, \ldots, U_n}{U_1, \ldots, U_n}.
   \end{align}
\end{widetext}
The expression $\cancel{U_p}$ denotes excluding the column
corresponding to this index from the determinant. It will also be useful to
define the kernel at each order $K_n(Z, Y)$, so that
$K(Z,Y) = \sum_{n=1}^\infty K_n(Z,Y) U^n$.

In this paper, we focus on the retarded Green function
\begin{equation}
   G^R(t,t') = -i \theta(t - t') \expval{ \qty{ c^\phdag_{0,\uparrow}(t), c_{0,\uparrow}^\dag(t')} },
\end{equation}
where $\expval{\hdots}$ represents the quantum average and
$\qty{A, B}$ the anticommutator between $A$ and
$B$. We aim to compute the perturbation series
\begin{equation}
   G^R(t,t') = \sum_{n=0}^{+\infty} G_n^R(t,t') U^n.
\end{equation}
As a consequence, we only need to consider $0 \leq t \leq t' = t_M$. Throughout,
we fix $t_M = 200/\Gamma$.

\subsection{Quasi-Monte Carlo integration: two algorithms}
\label{sec:QQMC}

In this section, we discuss how to use low-discrepancy sequences to compute the
integrals in \eqnref{eq:kernel}, which define the kernel. We build on the
work of Ref.~\onlinecite{Macek2020} where QQMC was used to compute single
quantities, such as the charge on the impurity or the current flowing through
it, and we start by briefly summarizing the approach. For a more detailed
explanation of QQMC, we refer to Ref.~\onlinecite{Macek2020}.

QQMC is a deterministic method which evaluates the perturbation theory
integrals such as in \eqnref{eq:integral_gf} at a given expansion order
$n$. Let us write such integrals schematically as
\begin{equation}
   I_n = \int_{[0, t_M]^n} \dd[n]{\vec{u}} \phi_n(\vec{u}).
\end{equation}
where $\phi_n$ is a generic scalar function. We wish to evaluate
this expression using points in a low-discrepancy sequence
$\vec{x}_i \in [0, 1]^n$. To modulate the density of samples with the amplitude
of the integrand, the integral is warped \cite{Macek2020}, i.e. a change
of variable $\vec{u} \rightarrow \vec{x}$ is applied to the integral. The warped
integral takes the form
\begin{equation}
   I_n = \int_{[0,1]^n} \dd[n]{\vec{x}} \phi_n[\vec{u}(\vec{x})]
   \qty| \frac{\partial \vec{u}}{\partial \vec{x}} |.
\end{equation}
This operation is meant to make the integrand as flat and smooth as possible in
the new variables, while allowing the transformation of the sampling sequence
$\vec{x}_i \rightarrow \vec{u}(\vec{x}_i)$ at low computational cost. The motivation is that,
unlike Monte Carlo, the rate of convergence of quasi-Monte Carlo improves with
the smoothness of the integrand. The change of variable is derived from a
\emph{model function} $p_n(\vec{u})$ that approximates the integrand
amplitude $|\phi_n(\vec{u})|$ and is similar to a re-weighting function in
Monte Carlo methods. The change of variable is defined implicitly by the model
function \cite{Macek2020}, such that
\begin{equation}
   \qty| \frac{\partial \vec{u}}{\partial \vec{x}} | = \frac{\mathcal{C}_n}{p_n(\vec{u})}, \quad
   \mathcal{C}_n = \int \dd[n]{\vec{x}}
   p_n(\vec{u}).
\end{equation}
Therefore, the integral reads
\begin{equation}
   I_n = \mathcal{C}_n \int \dd[n]{\vec{x}}
   \frac{\phi_n[\vec{u}(\vec{x})]}{p_n[\vec{u}(\vec{x})]},
\end{equation}
which is evaluated using the first $M$ elements of a
low-discrepancy sequence $\{\vec{x}_i\}$
\begin{equation}
   I_n \approx \frac{\mathcal{C}_n}{M} \sum_{i=1}^M
   \frac{\phi_n[
      \vec{u}(\vec{x}_i)]}{p_n[\vec{u}(\vec{x}_i)]}.
\end{equation}
For fast convergence, the model function must capture both the overall
structure and asymptotic decay of the integrand \cite{Macek2020}.
Consequently, the choice of the model function depends on the parameters of the
model, the perturbation order, and the quantity to compute. This choice is made
automatically by a projection algorithm outlined in
Ref.~\onlinecite{Macek2020}, related to the VEGAS algorithms
\cite{Lepage_1978,Lepage_1980}. We refine this procedure in the present work, as
described in details in Sec.~\ref{sec:projection_warper}.
In this work we use a Sobol' sequence as the low-discrepancy sequence. For
error estimation, we use the standard technique of randomized quasi-Monte Carlo
\cite{Dick_2013,Nuyens_1308,Dick_Pillichshammer_2010,Lecuyer_2018}: we compute separate results from 10 randomized Sobol'
sequences, and take the standard deviation as an error estimate.

We developed two algorithms to compute the kernel with a low-discrepancy
sequence, which we now describe.

\subsubsection{Single frequency algorithm}
\label{sec:single-frequency}
One way to apply QQMC to calculate the kernel is to compute one frequency
$\omega$ at a time. The retarded Green function in the
stationary regime reads \cite{Bertrand_1903_kernel}
\begin{equation}
   G^R(\omega) = g^R(\omega) + K^A(\omega)^\dagger g^R(\omega),
\end{equation}
where the advanced kernel in the stationary limit is
\begin{equation}
   K^A(t) = \lim_{t' \rightarrow +\infty} [K^{00}(t + t', t') - K^{10}(t + t', t')].
\end{equation}
Using the definition of $K$, \eqnref{eq:kernel}, the
perturbation series for $K^A(\omega)$ at order
$n$ reads
\begin{multline}\label{eq:single-freq}
   K_n^A(\omega) = \lim_{t' \rightarrow +\infty} \int \dd{u_1} \ldots
   \dd{u_n} \\ \sum_{p=1}^{n} \sum_{a_p} (-1)^{a_p}
   W^n_p(Y, \vec{u}, a_p) e^{i\omega (u_p - t')}.
\end{multline}
The integral is taken on the $[0, t']^n$ hypercube, and a large
value of $t'$ approximates the stationary regime.
Equation~\eqref{eq:single-freq} defines, for given $\omega$ and
$t'$, a standard $n$-dimensional
integral. It can be evaluated using any high-dimensional integration technique,
in particular QQMC. We will refer to this algorithm, using QQMC, as the
\emph{single frequency} method.

\subsubsection{Full kernel algorithm}
In the single frequency technique, computation has to be repeated for different
frequencies. This may become a drawback if one is interested in a high
resolution spectrum or a large range of frequencies. This is why we consider a
second algorithm, referred as \emph{full kernel} method, where the whole
time-dependent kernel is computed at once. The idea is reminiscent of the
original usage of the kernel in Ref.~\onlinecite{Bertrand_1903_kernel}, but using
quasi-Monte Carlo and the warping technique of QQMC. We integrate
\eqnref{eq:kernel} for many values of $Z$ using a
single Sobol' sequence of vectors $\vec{u}$. Because of the delta
function $\delta(Z, U_p)$, each vector $\vec{u}$ provides
values of $K_n$ for $2n$ different points
$Z$ (both Keldysh indices at each $u_p$).
There is therefore no unique scalar integrand.

How does the notion of warping generalize to such integrals? Although the
integrand is not a conventional scalar integral, we still need to provide a
unique model function. To do so, we consider at each order
$n$ a weight function
\begin{equation}
   \label{eq:weight-function}
   W_n(\vec{u}) = \sum_{p=1}^{n} \sum_{a_p}
   \qty| W^n_p(Y, \vec{u}, a_p) |,
\end{equation}
which is independent of $Z$ (note that
$Y$ is fixed). We note that $W_n$ was
already used as the weight function in the Monte Carlo\ignorespaces
\footnote{The algorithm of Ref.~\onlinecite{Bertrand_1903_kernel} actually sampled different orders
$n$ in a single Markov chain and used a more generalized
weight.} of Ref.~\onlinecite{Bertrand_1903_kernel}. Since
$W_n$ is computationally expensive, the QQMC model function
$p_n$ is built as a low-rank approximation to it. Because
this is independent of $Z$, we expect it to be less
efficient than model functions optimized for each fixed value of
$Z$.

With this model function, the warped integral
\begin{equation}
   K_n(Z,Y) = \mathcal{C}_n \int \dd[n]{\vec{x}} \frac{f_n[Z,Y,
   \vec{u}(\vec{x})]}{p_n[\vec{u}(\vec{x})]},
\end{equation}
can be efficiently evaluated using a low-discrepancy sequence.
As already mentioned, each sample $\vec{x}_i$ provides
contributions to $K_n$ to $2n$ different
values of $Z$. In practice, these are binned into a
histogram on a fine time mesh \cite{Bertrand_1903_kernel} (we use $50\,000$ bins).

\subsection{Projection-based model function}
\label{sec:projection_warper}

We now turn to the choice of a model function $p_n$, which is
crucial to the quality of the warping and the success of the quasi-Monte Carlo
approach. We advise readers who prefer to see results before methodological
details to jump directly to Sec.~\ref{sec:convergence}, and read this section
later.

To obtain a good approximation of $W_n$ at a reasonable
computing cost, we use the projection technique described in
Ref.~\onlinecite{Macek2020} (Sec.~IX.B of Supplementary Material). Here we have made improvements, which allow it to
be more robust and automatic. Among them is using the model function of one
order as a starting point for the next order, thus reducing considerably the
effort of building high order model functions. Our procedure forms a good
warping for each order with only one manually fixed parameter. In the following
we summarize the projection technique, describe these improvements and finally
compares the model function created to the weight function it approximates.

The warping procedure is a succession of three changes of
variable~\cite{Macek2020}
\begin{equation}
   \vec{u} \rightarrow \vec{v} \rightarrow
   \vec{w} \rightarrow \vec{l}.
\end{equation}
The components of $\vec{u}$ can be assumed sorted so that
$0 < u_n < \ldots < u_1 < u_0 = t'$. The first change of variable is defined simply by
$v_i = u_{i-1} - u_i > 0$. It maps the $\vec{u}$-hypercube
$[0, t']^n$ into a simplex included in the
$\vec{v}$-hypercube $[0, t']^n$. During the
integration the full $\vec{v}$-hypercube is sampled, but the
contribution of points that lie outside the simplex is set to zero in order to
respect the integration domain~\cite{Macek2020}. In practice, only a few
percents of points are rejected this way. Indeed, the distribution of points in
the $\vec{v}$-hypercube is not uniform, but defined by the two
next changes of variable.

The second change of variable $\vec{v} \rightarrow \vec{w}$ is defined by a model
function \cite{Macek2020}
\begin{equation}
   \label{eq:pre-model-func}
   p^{\rm pre}_n(\vec{v}) =
   \prod_{i=1}^n h^{\rm pre}_{i,n}(v_i),
\end{equation}
which approximates roughly the weight function $W_n$ in the
$\vec{v}$-space. The $\vec{v}$-hypercube
$[0, t']^n$ is mapped onto the $\vec{w}$-hypercube
$[0, 1]^n$ via
\begin{equation}
   w_i = \int_0^{v_i} \dd{y}
   h_{i, n}^{\rm pre}(y).
\end{equation}
This model function aims at roughly capturing the long time tails of the weight
function and acts as an importance sampling method for the construction of the
last change of variable. For this reason we call it preliminary model function.
The actual choice of $h^{\rm pre}_{i,n}$ is detailed later in this section.

The last change of variable $\vec{w} \rightarrow \vec{l}$ is defined by another model
function
\begin{equation}
   \label{eq:proj-model-func}
   p^{\rm proj}_n(\vec{w}) =
   \prod_{i=1}^n h^{\rm proj}_{i,n}(w_i),
\end{equation}
which approximates the weight function $W_n$ in the
$\vec{w}$-space. This time however, $h^{\rm proj}_{i,n}$ are
constructed by projecting $W_n(\vec{w})$ on each axis of the
$\vec{w}$-space
\begin{equation}
   \label{eq:projection}
   h^{\rm proj}_{i,n}(y) = \int_{[0,1]^n}
   \dd[n]{\vec{w}} W_n(\vec{w})
   \delta(w_i - y).
\end{equation}
This construction is done by sampling the $\vec{w}$-hypercube
with a Sobol' sequence, propagating the samples in the
$\vec{u}$-space where $W_n$ can be computed
with \eqnref{eq:weight-function}, and projecting the resulting values into
$n$ different histograms. Each histogram is made of
$N_{\rm bin} = 500$ bins. Because of the forms of the two model functions,
Eqs.\eqref{eq:pre-model-func} and~\eqref{eq:proj-model-func}, the composed change of
variable $\vec{v} \rightarrow \vec{l}$ is described by a model function of the same
form \cite{Macek2020}
\begin{gather}
   \label{eq:full-model-func}
   p_n(\vec{v}) = \prod_{i=1}^n
   h_{i,n}(v_i), \\ h_{i,n}(v_i) =
   h^{\rm pre}_{i,n}(v_i) \ h^{\rm proj}_{i,n}(w_i).
\end{gather}

We are only interested in having a good approximation of
$W_n(\vec{w})$ in the part of the $\vec{w}$-space which
is ultimately used in the integration. Nevertheless, the projection
\eqnref{eq:projection} takes the whole $\vec{w}$-hypercube
$[0,1]^n$ into account, so the values of $W_n$
outside the integration domain have an effect on the final warping. When
building the warping, these points are evaluated and not set to zero.

Given that high accuracy is not necessary in this step, and that there is no
sign problem, we use only a few $10^6$ evaluations to perform
the projection. Therefore, the histograms in which the projections are stored
approximate the $h^{\rm proj}_{i,n}$ at coordinates $y_j = (j - 1/2) / N_{\rm bin}$
($j=1, \ldots, N_{\rm bin}$) with the addition of noise. To reduce this noise, we
smooth them using a local linear regression. This is less biased than the
kernel smoothing used in Ref.~\onlinecite{Macek2020}, in particular near the
boundaries where the majority of the $\vec{u}$-hypercube is
mapped into. However, it can yield negative values, even when the input is
strictly positive. We therefore do the linear regression in log space, to
ensure the result to be strictly positive. To be precise, we apply:
\begin{equation}
   h^{\rm proj}_{i,n}(y_{j_0}) \rightarrow
   \exp(a_{j_0} \ y_{j_0} + b_{j_0})
\end{equation}
where $a_{j_0}$ and $b_{j_0}$ are the slope and
intercept of the weighted linear regression of values $\log\left(h^{\rm proj}_{i,n}(y_j)\right)$
at coordinates $y_j$ with weight $\exp(-(y_j - y_{j_0})^2 / \lambda^2)$, for
$j = 1, \ldots, N_{\rm bin}$. We use $\lambda = 0.01$. Empty bins are ignored
in the linear regression, although a proper choice of $h^{\rm pre}_{i,n}$
and enough sampling should reduce chances that it happens.

Finally, it remains to define the $h^{\rm pre}_{i,n}$. These should be fast
to compute and capture roughly the long time tails of the weight function ---
which take the largest part of the integration space. As we generally compute
order by order, we choose to reuse the model function of order
$n-1$ to define the preliminary model function of order
$n$. Namely, we use:
\begin{align}
   \label{eq:recursion_first_order}
   h^{\rm pre}_{1,1}(v) ={} & \frac{1}{(1 + v)},
   \\
   \label{eq:recursion_main}
   h^{\rm pre}_{i,n}(v) ={} & h_{i,n-1}(v) \qfor 1 \le i < n,
   \\
   \label{eq:recursion_last}
   h^{\rm pre}_{n,n}(v) ={} & h^{\rm pre}_{n-1,n}(v).
\end{align}
We justify \eqnref{eq:recursion_main} by the observation that the projections of
the weight function on a given axis $i$ are similar between
adjacent orders. To complete the model, we duplicate the function at
$i=n-1$ for $i=n$ (\eqnref{eq:recursion_last}).
Indeed, we observed that at any order the projection onto the last axes look
very similar. As the starting point of this recursive definition,
$h^{\rm pre}_{1,1}$, we chose an arbitrary analytic function -- here an
inverse function (\eqnref{eq:recursion_first_order}).

\begin{figure}[t]
   \centering
   \includegraphics[width=\columnwidth]{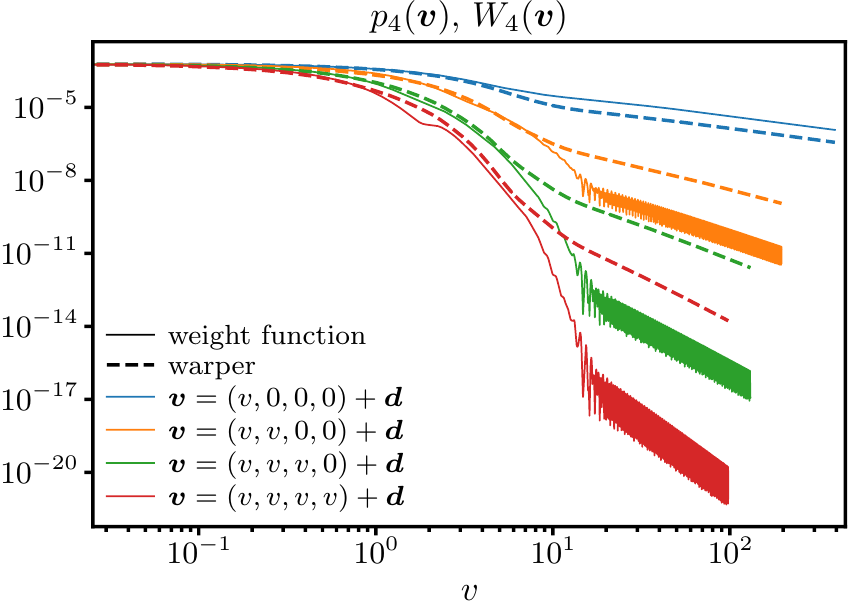}
   \caption{
      \label{fig:warper}
      The weight function $W_n$ at order $n=4$ (plain
      lines) compared to its model function $p_4$ built by the
      projection technique (dashed lines), in the $E_d = \Gamma$ case. These functions are shown along several
      lines in the $\vec{v}$-space, parameterized by $v$ and starting at the point
      $\boldsymbol{d} = (2, 2, 2, 2)$. Despite the simplistic form of the model function
      \eqnref{eq:full-model-func}, it gives a good approximation in the low $v$ region and
      captures precisely the power law scaling of the long time tails.
   }
\end{figure}

The model function obtained with this method at order $n=4$
is compared to the weight function in Fig.~\ref{fig:warper}, in the
$E_d = \Gamma$ case. Values along different lines in
$\vec{v}$-space are displayed in different colors. Values close
to $\vec{v} = \vec{0}$ are well approximated, but a large difference
appears in the large $\| \vec{v} \|$ tails. However, it is remarkable,
and very important for long time calculations, that the model function captures
the correct power law scaling of these tails. The difference can be explained
from the simplicity of the model function \eqnref{eq:full-model-func}, which does
not capture the full complexity of the weight function.

\subsection{Results}
\label{sec:convergence}

\begin{figure*}[t]
   \centering
   \includegraphics[width=\linewidth]{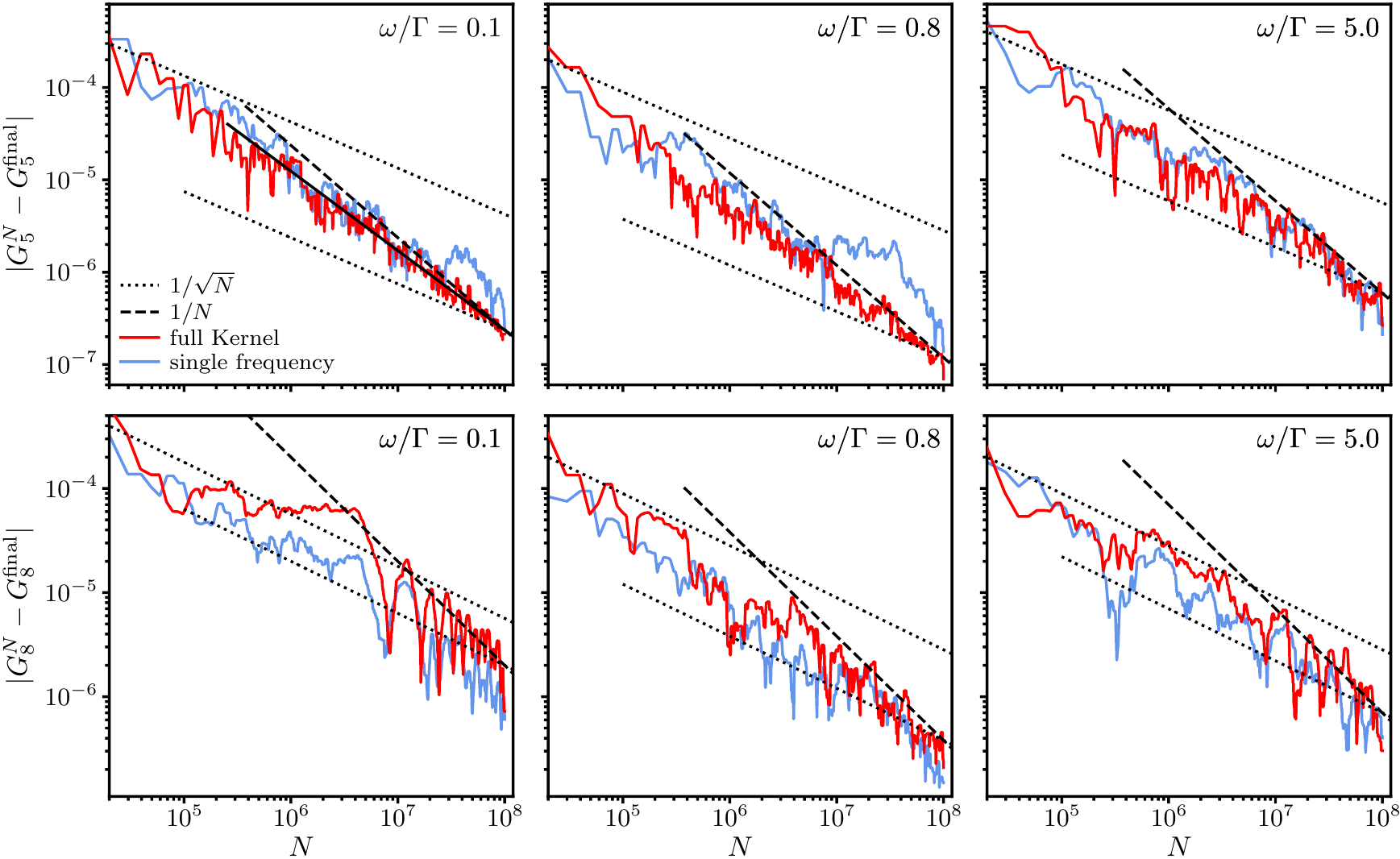}
   \caption{
   \label{fig:convergence}
   Convergence of the absolute error of the coefficients $G^R_n(\omega)$
   calculated by QQMC with increasing number of function evaluations
   $N$. The full kernel calculation (red lines) is compared to
   the single frequency integration \eqnref{eq:single-freq} (blue lines). Three frequencies
   $\omega/\Gamma = 0.1, 0.8$ and $5.0$ (columns) are shown at order
   $n=5$ (top row) and $n=8$ (bottom row). The
   model is the asymmetric one $E_d=\Gamma$. The scaling of the error is
   systematically better than or similar to $1/\sqrt{N}$ (dotted black
   lines), and is close to $1/N$ (dashed black lines) at small
   frequencies, reaching $1/N^{0.86}$ in the best case (black plain line
   in top left panel). For readability we show an upper bound to the error, see
   main text for details, and for the error estimation method. }
\end{figure*}

Let us now apply QQMC to calculate the coefficients of the retarded Green
function $G^{R}_n(\omega)$ of \eqnref{eq:Htot} in the perturbative
expansion.

Figure~\ref{fig:convergence} shows the convergence of the coefficients at
order $n=5,8$ for three different frequencies
$\omega / \Gamma =  0.1, 0.8$ and $5.0$. Specifically, we show the
evolution of the absolute error with the number of function evaluations
$N$. Each panel shows the two different methods outlined in
the previous section -- fixed single frequencies (blue curves) and full kernel
method (red curves). The errors are estimated by taking the deviation of the
value obtained for $G^R_n(\omega)$ after $N$
samples ($G^N_n$ in the figure) from the final value at
$N=10^8$. The final value $G^{\textrm{final}}_n$ is taken to be
the average of the last $10^6$ values. To improve readability,
we show an upper bound to the error consisting of the maximum of a moving
window around $N$ of fixed relative size
($4\%$ of $N$).

We focus first on the full kernel method. We observe that the convergence is
systematically better than $1/\sqrt{N}$ (dotted line). It shows the
best convergence in the top left panels ($\omega/\Gamma = 0.1$ and
$0.8$, $n=5$), where we observe convergence
with a clear power law $1/N^{0.86}$ (black line). In the other
panels ($n=8$ or $\omega/\Gamma = 5$), the convergence is
slower, but never worse than $1/\sqrt{N}$ (dotted lines). The
slowdown at large order is expected, as it is more difficult for the model
function to capture the details of the integrand at high dimension. Notice that
at order $n=8$, the convergence is characterized by a slower
rate for $N < 3\cdot 10^6$ than for $N > 3\cdot 10^6$. This
separation in two regimes was already observed in Ref.~\onlinecite{Macek2020}.

As the single frequency method computes a single integral, it could be expected
that the distribution of samples chosen by the projection technique is more
adapted than the distribution used in the full kernel calculation. However, we
see no significant improvement in using the single frequency integration:
scalings are similar as well as absolute error values. The similarity between
both methods convergences show that the weight function
Eq.~\eqref{eq:weight-function} is an efficient distribution for computing all
frequencies at once, given that we approximate all distributions by a
projection-based model function.

\begin{figure}[t]
   \centering
   \includegraphics[width=\columnwidth]{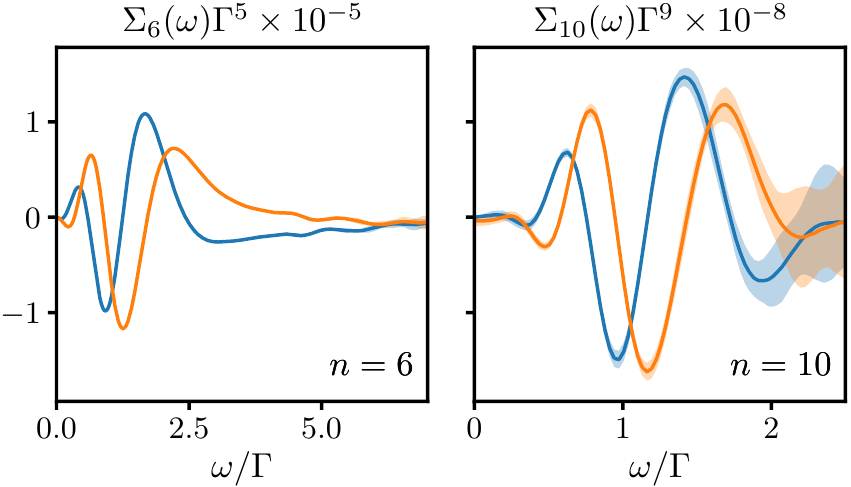}\\[2mm]
   \includegraphics[width=\columnwidth]{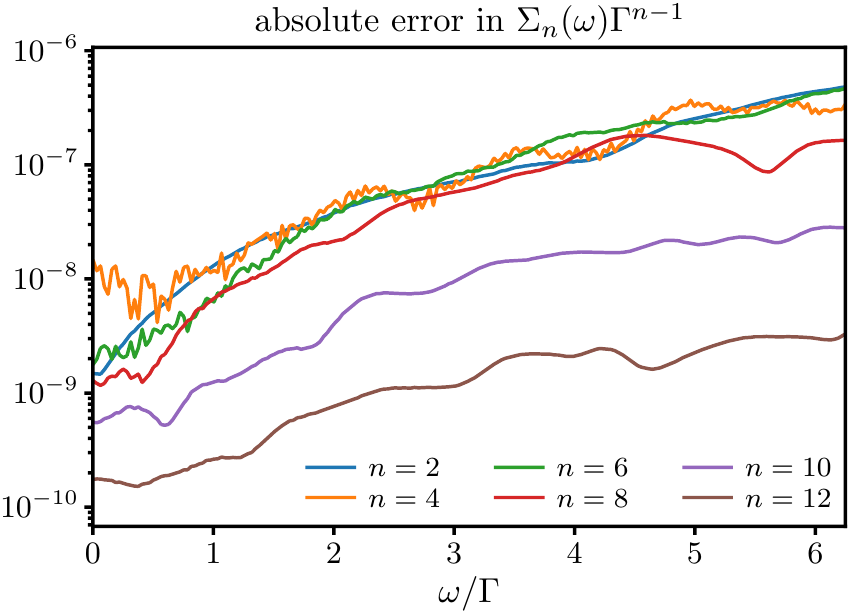}
   \caption{
      \label{fig:mcmc-vs-qqmc}
      Self-energy series coefficients and absolute error, using the full kernel method on the $E_d=0$ impurity model. Top
      panels: self energy coefficient at order 6 (left) and 10
      (right),
      blue is the real part, orange the imaginary part. Error bars are shown as a shaded area. Bottom panel:
      absolute error at different orders $n$. We used 10 shifted
      Sobol' sequences of $10^8$ samples, and the error is the standard deviation between their results.
   }
\end{figure}

We now study the error as a function of frequency for different orders. The
lower panel of Fig.~\ref{fig:mcmc-vs-qqmc} shows the absolute error in the
self-energy coefficients $\Sigma_n(\omega)$ using the full kernel method
and $N=10^9$ samples in the $E_d = 0$ case. The
self-energy is computed from the Green function series using Dyson's equation
$\Sigma(\omega) = g^R(\omega)^{-1} -
G^R(\omega)^{-1}$. The error is estimated here by taking the standard
deviation of the results of 10 randomized Sobol' sequences. We see that the
error is frequency dependent, with a minimum at $\omega = 0$
reaching as low as a few $10^{-10}$ at large orders. The full
kernel method is therefore well suited for low frequencies. The error does not
change significantly as the order increases until $n=8$, but
the coefficients decrease in absolute value by about an order of magnitude each
1--2 perturbation order (see Appendix~\ref{app:series}). Indeed, as the
integration dimension increases, the relative error of the integral
deteriorates. The top panels show some self-energy coefficients, with the error
indicated as a shaded area. Order 6 (left) is well resolved for all frequencies
with a significant signal. Order 10 (right) sees some deterioration due to the
higher integration dimension, but accuracy is still good on a large range of
frequencies. Only at high frequencies ($\omega > 2\Gamma$) does the
relative error become too large.

In brief, Ref.~\onlinecite{Macek2020} showed that evaluating at a low
discrepancy sequence of points makes it possible to compute single observables
with an error decreasing faster than $1/\sqrt{N}$. Here, we
establish that this also holds for dynamical quantities on a large range of
frequencies, computed altogether with a single sampling. We also show that the
kernel technique is well suited for that task, in particular at low
frequencies.

\section{Resummation with Padé approximants}
\label{sec:pade}

In the previous section, we described how to compute the frequency-dependent
perturbation series coefficients $G^R_n(\omega)$ of the Green function.
We showed that applying QQMC is efficient in a large frequency range, and we
observed convergence scalings that outperform the $1/\sqrt{N}$ of
conventional Monte Carlo. This was enabled by the automatic construction of a
simple and computationally cheap -- yet robust		    -- model function.
We can now resum these series to obtain physical quantities of interest. These
can be compared with other numerical methods and we will perform a detailed
comparison between QQMC and FTPS in later sections. In this section, we will
discuss the series resummation of the Green function at values of the
interaction $U$ beyond the radius of convergence.

There are several ways to perform resummation, with different performance
characteristics. In Ref.~\onlinecite{Bertrand_1903_series}, some of us designed a robust
and general-purpose resummation technique based on conformal transforms; these
were benchmarked on the Green function for an Anderson impurity system similar
to the model of Section~\ref{sec:model}. Instead of repeating that
approach, we will perform resummation using Padé approximants, which are
well-established method for analytical continuation~\cite{Hunter1979, Rossi_2001,
Pavlyukh2017, Simkovic2019}.
Here, we only provide a summary of relevant aspects and refer  to the
literature \cite{baker1996pade} for a detailed exposition. Note that in our
analysis, we resum the perturbation series at each frequency
$\omega$ independently; we will therefore generally suppress the
$\omega$ dependence in the notation of this section.

The Padé approximant of type $[l / m]$ of a series is the unique
rational function $P/Q$, with $P$ of
degree at most $l$ and $Q$ of degree at
most $m$, whose Taylor expansion at the origin matches the
series up to the highest order possible \cite{baker1996pade}. When the
series represents a function $f(U)$, such an approximant
respects
\begin{equation}
   P(U) - f(U) Q(U) = O(U^{l+m+1})
\end{equation}
and generalizes the truncated Taylor series as an approximation of
$f$ in the $U$ complex plane. However,
unlike Taylor series, Padé approximants can capture the locations of poles and
can be accurate beyond the convergence radius \cite{baker1996pade}.
The choice of $l$ and $m$ is important
for obtaining a good approximant. As the infinite $U$ limit
is known $G^R(\omega) \rightarrow 0$, for $\omega \neq 0$, we impose the
restriction $m \ge l$.
For the purposes of this paper, we choose the simplest Padé that was
sufficiently well matched with FTPS results presented below. Trying to find
this Padé in a way that is independent of FTPS results is an involved process
which we did not pursue here.

A recurrent issue with rational approximants is the occurrence of so-called
\emph{defects} or Froissart doublets \cite{baker1996pade,Stahl1998}. They are
produced when $P/Q$ is close to a singular Padé approximant,
in which a root of $P$ equals one of
$Q$. The defect manifests as a localized zero-pole pair,
which produces dramatic variations when the approximant is evaluated close-by,
but has vanishing influence at long distance.

While defects are inevitable elements of Padé approximants, their locations are
sensitive to the exact values of the coefficients $G^R_n$, so
that noise may move them closer to or away from the point
$U$ of interest. This causes extreme variance in the
resummation of the series as $\omega$ varies and as the noise is
changed (see Appendix~\ref{app:elim_defects} for an example). Such
instabilities are evidence of the presence of a defect, and should be
eliminated.

Some defects -- but not all of them -- are sensitive to the noise introduced by
the integration method in both QQMC and traditional Monte Carlo. These may be
statistically removed as follows. If each coefficient $G^R_n$
is known within an error bar $\delta G^R_n$, we assume it can be
represented by a random variable following a normal probability law centered on
$G^R_n$ and of standard deviation $\delta G^R_n$; we
ignore potential correlations between orders. By sampling the series
coefficients from this distribution (in practice we take 100 samples), we
obtain as many Padé approximants, which we evaluate at the target
$U$. This gives a population of resummed values at
$U$, from which we take the median of the real or imaginary
part as the final resummed result. The 15th and 85th percentiles are taken as
the propagation of the coefficients error bar (these percentiles correspond to
one sigma in the normal distribution). Note that these do not contain the error
made by the resummation itself.

The population of resummed values do not form a Gaussian distribution, as would
have been expected if using conformal transforms. Padé resummation is a
non-linear process and the presence of defects close to the target
$U$ brings outliers. For these reasons, median and
percentiles are preferred over average and standard deviation. Nonetheless, the
error bar obtained from percentiles still reflects the increased sensitivity of
the Padé approximant in the series coefficients, in the presence of a defect
near the target $U$.

Note that other defects may exist which are not susceptible to variations of
the coefficients within error margins. These cannot be detected or eliminated
statistically. Nevertheless, all the results in this article are resummed with
a choice of $l$ and $m$  which shows no
sign of such a defect in the vicinity of the target $U$.

Finally, we remark that by resampling the coefficients, we neglected the
correlations between them. This makes it difficult to propagate error bars back
in time domain after the Padé resummation.

\section{FTPS}
\label{sec:FTPS}

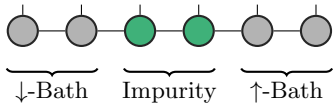
\begin{figure}[t]
   \centering
   \begin{tikzpicture}
    
   \definecolor{DanBlue}{RGB}{66,133,244}
   \definecolor{DanRed}{RGB}{219,68,55}
   \definecolor{DanGreen}{RGB}{15,157,88}
   \definecolor{DanYellow}{RGB}{244,160,0}

    \tikzstyle{old}=[	circle, thick, minimum size=0.4cm,
    				draw=black!85, fill=DanBlue!80, inner sep=0pt,  text width=4mm]

     \tikzstyle{new}=[	circle, thick, minimum size=0.4cm,
    				draw=black!85, fill=DanRed!80, inner sep=0pt, text width=4mm]
    
   \tikzstyle{act}=[	circle, thick, minimum size=0.4cm,
   				draw=black!85, fill=DanYellow!80, inner sep=0pt, text width=4mm]
    
   \tikzstyle{Imp}=[	circle, thick, minimum size=0.4cm, draw=black!85,
				fill=DanGreen!80, inner sep=0pt, text width=4mm]
				
   \tikzstyle{Bath}=[	circle, thick, minimum size=0.4cm, draw=black!85,
				fill=black!30, inner sep=0pt, text width=4mm]
    
    	\matrix (m2) [matrix of nodes, column sep=10, row sep = 8] {
       		|[Bath]| & |[Bath]| & |[Imp]| & |[Imp]| & |[Bath]| & |[Bath]|  \\
    	};
    
    	\foreach \x/\y in {1/2,2/3,3/4,4/5,5/6} \draw (m2-1-\x) -- (m2-1-\y);
	\foreach \x in {1,2,3,4,5,6} \draw (m2-1-\x) -- ($(m2-1-\x)+(0,0.35)$);

	\draw [ 	thick, decoration={ brace, mirror, raise=0.5cm },
    			decorate]  (m2-1-1.west) -- (m2-1-2.east) %
		node [pos=0.5,anchor=north,yshift=-0.55cm] {$\downarrow$-Bath}; 
	\draw [ 	thick, decoration={ brace, mirror, raise=0.5cm },
    			decorate]  (m2-1-3.west) -- (m2-1-4.east) %
		node [pos=0.5,anchor=north,yshift=-0.55cm] {Impurity}; 
		
	\draw [ 	thick, decoration={ brace, mirror, raise=0.5cm },
    			decorate]  (m2-1-5.west) -- (m2-1-6.east) %
		node [pos=0.5,anchor=north,yshift=-0.55cm] {$\uparrow$-Bath};

\end{tikzpicture}
   \caption{
      \label{fig:MPS}
      Depiction of the MPS used for a impurity model with a small bath consisting of
      two sites. Note that the main idea of FTPS is to separate the spin-degrees of
      freedom (orbital-degrees of freedom in the multi-orbital case). Note that the
      sites one the left and on the right of the impurity does not correspond to the
      left or right lead as in Eq.~\ref{eq:H0} but represent the two
      different spin-degrees of freedom of the equilibrium bath. }
\end{figure}

In the recent years, Tensor Network (TN) methods -- especially those based on
Matrix Product States (MPS) -- have been extensively applied to impurity
problems~\cite{JeckelmannDDMRG, Hallberg_DMRGImpuritySolver,  GanahlMPSImpSolver,
WolfStarGeometry, WolfChebychev, WolfImagTime,  BauernfeindFTPSorig, LindenSrUO}. They allow for a
systematically improvable
representation of the impurity problem at all energy scales as well as
well-developed approaches for real-time and imaginary time evolution. Here, we
benchmark our QQMC results with those obtained from the Fork Tensor Product
States (FTPS) impurity solver~\cite{BauernfeindFTPSorig}. FTPS is a TN that is
especially suited for multi-orbital impurity problems and it can be used to
compute the impurity Green function on the real-frequency axis. This is
achieved by a Density Matrix Renormalization Group
(DMRG)~\cite{WhiteDMRGorig, Schollwoeck_1008} calculation for the ground state followed by a
time evolution in real time. For the single orbital model studied in this work,
FTPS reduces to a Matrix Product State (MPS) as shown in
Fig.~\ref{fig:MPS} but we keep the term FTPS since certain details of
the algorithm used to solve the impurity problem differ from standard MPS
algorithms (see App.~\ref{app:TDVP}).

A comparison of QQMC
with FTPS is a fruitful endeavor since the approximations made in the two
algorithms are drastically different. FTPS is non-perturbative and its accuracy
can be systematically improved. However, it is wave-function based, and
therefore solves a discretized version of the Anderson impurity model: a large
but finite bath which consists of $N_b$ sites is used to
represent the hybridization $\Delta^R(\omega)$ on a regular energy grid.
The effect of such a discretization is that there exists a time
$t_{\rm max}^{\rm FTPS} = \pi N_b/D$ after which the Green function shows finite size
effects. Before that ($t<t_{\rm max}^{\rm FTPS}$), finite size effects are very
small and the result behaves like the Green function of a model with the
continuous bath. In this work we use $N_b = 409$, so that
$t_{\rm max}^{\rm FTPS} \approx 112 / \Gamma$ in our parameters.

FTPS performs the computation in the so-called star geometry representation of
the bath~\cite{Anderson1961, BullaNRG, Caffarel1994,
WolfStarGeometry}. In this representation, each bath degree of
freedom is coupled directly to the impurity. Although this introduces
long-range hopping terms in the Hamiltonian, this representation is superior
for TN methods as it turns out~\cite{WolfStarGeometry}.

We use FTPS to calculate the equilibrium zero-temperature retarded Green
function $G^{R}(t) = -i \theta(t)
\bra{\psi_0} \{ c^\phdag_{0,\uparrow}(t),
c_{0,\uparrow}^\dag \} \ket{\psi_0}$ in real-time. To compute
$G^R(t)$, FTPS first computes the ground state
$\ket{\psi_0}$ using the Density Matrix Renormalization Group (DMRG),
and time-evolves the states with an additional impurity electron/hole using the
Time Dependent Variational Principle (TDVP) technique~\cite{HaegemanUnifyingTevo}
in its two-site variant. TDVP can be considered as a set of coupled
differential equations which are usually integrated in a certain order to
obtain an algorithm very similar to DMRG~\cite{HaegemanUnifyingTevo}. FTPS uses a
different integration order as discussed in App.~\ref{app:TDVP}.

Using this approach, we perform the time evolution up to
$t=40/\Gamma$ using a time step of $\Delta t = 0.05/\Gamma$. To
account for the finite maximum time we Fourier transform with a modified kernel
$e^{i\omega t - \eta |t|}$, which generates a Lorentzian broadening of width
$\eta$ in energy space. For the models studied in this work,
the broadening is set to $0$ for spectral functions, since
the Green functions in time decay rapidly enough (see
Fig.~\ref{fig:bench-time}). For the impurity self-energy, on the other hand,
broadening is necessary because it is calculated from Dyson's equation, which
implies the non-interacting Green function calculated from the finite sized
bath. This means that the non-interacting Green function consists of Dirac
deltas with energy difference $\Delta \epsilon = \frac{2D}{N_b}$ which turns out to be
rather large in our parameters: $\Delta \epsilon \approx 0.06 \Gamma$ making some form of
extrapolation necessary.

To obtain the $\eta \to 0$ self-energy, we calculate it for various
broadenings $\eta$ and extrapolate each frequency point
towards $\eta \to 0$ using a 4th order polynomial
regression\footnote{For this, every term in the Dyson equation needs to be evaluated with the same value of
$\eta$ including the interacting Green function. To actually
perform the extrapolation, the $\eta$-values we use are ten values between 0.05 and 0.15.}. We checked that this approach is consistent
with the self energy obtained from the $\eta=0$ interacting
Green function and the continuous non-interacting Green function. The latter
yields worse self energies though, because Friedel oscillations that are barely
visible in the interacting Green function are enhanced by the inversion in the
Dyson equation.

The tensor network approximation used a truncated weight of
$10^{-12}$ (sum of all squared discarded Schmidt values) and the
maximal bond-dimensions were restricted to 300 for the link connecting the two
impurity degrees of freedom and 200 for all other links. We checked that the
results are converged with respect to larger bond dimensions and that they are
converged in the time step $\Delta t$.

\section{Comparing QQMC with FTPS}
\label{sec:comparison}

We now compare the results from the full kernel QQMC with those from FTPS on
the Anderson impurity model in the Kondo regime. In this section, we show that
the Green function perturbation series -- calculated by QQMC -- can produce
results that match FTPS when resummed with a simple Padé approximant.

For each of the two cases, $E_d = 0$ and $E_d = \Gamma$,
the retarded Green function has been computed up to order
$n=12$, and resummed in the frequency domain, as discussed
above. We use $N = 10^9$ function evaluations at each order;
previous calculations some of the authors made using a Markov chain Monte
Carlo~\cite{Bertrand_1903_series} obtained less accurate results with 30 times more
samples.

\begin{figure}[t]
   \centering
   \includegraphics[width=\columnwidth]{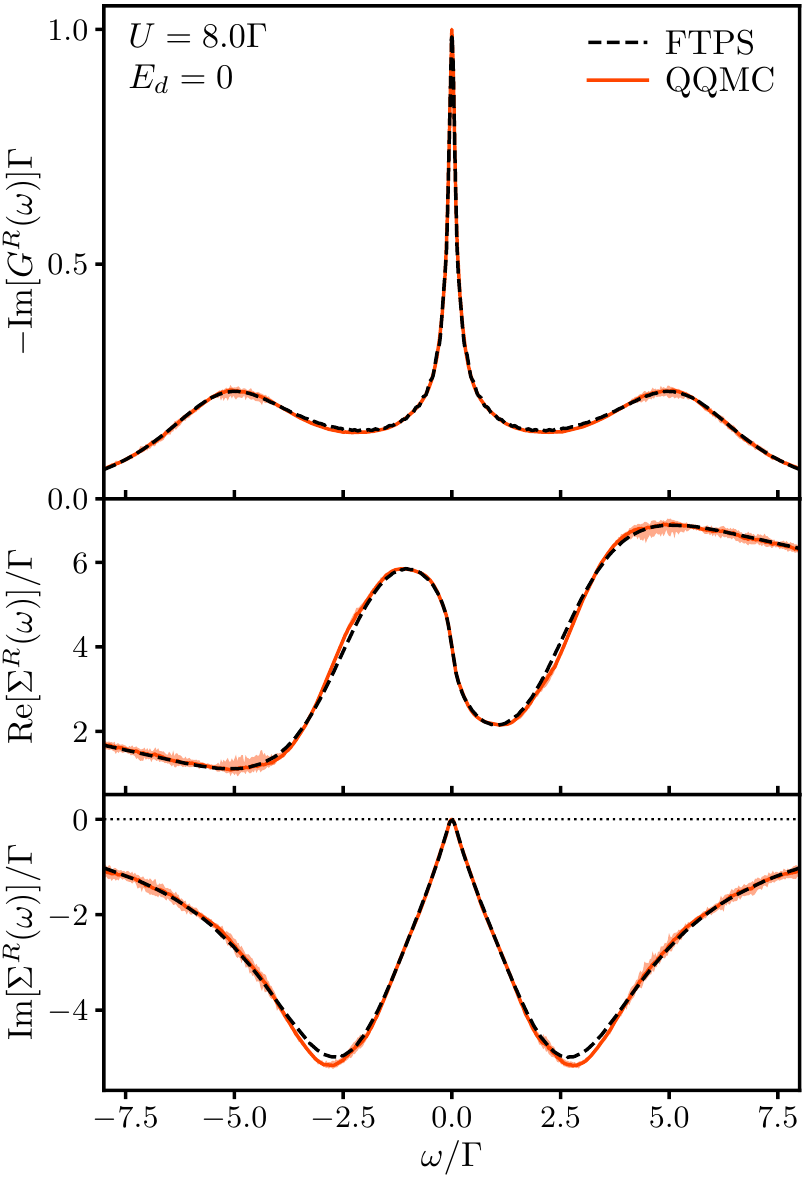}
   \caption{
      \label{fig:bench-sym}
      Comparison between QQMC result after resummation (plain line) and FTPS (dashed
      line), at $U=8\Gamma$ in the symmetric case $E_d = 0$. Top
      panel: spectral function on the dot. Middle and bottom panels: real and
      imaginary parts of the associated self-energy. Shaded areas represent QQMC
      integration error estimate only (see main text). QQMC used
      $N=10^9$ function evaluations at each order.
   }
\end{figure}

\subsection{Symmetric model in frequency}
Figure~\ref{fig:bench-sym} shows results for the symmetric model
($E_d = 0$) at $U=8\Gamma$, deep in the Kondo regime.
In this model, due to particle-hole symmetry, the retarded Green function
depends on $U^2$ instead of $U$. Its
series is resummed using the $[2/4]$ Padé approximant (in the
$U^2$ variable) at low frequencies $|\omega| < 2\Gamma$,
and $[2/3]$ at high frequencies. The high frequencies series
decreases faster with order, so that high order coefficients are not resolved,
and an approximant of lower rank is more accurate. The transition between the
two Padés is progressive over a range $0.25\Gamma$.

The calculation of the coefficients $G_n^R(\omega)$ is subject to an
error caused by the QQMC integration method, estimated as explained in
Sec.~\ref{sec:convergence}. In addition, the resummation of the series
produces another error, which is difficult to estimate as it is linked to
several factors (choice of Padé rank, finite perturbation series, defects). The
error bars in Fig.~\ref{fig:bench-sym} (shaded area) reflects the first error, propagated
through the resummation, as explained in Sec.~\ref{sec:pade}. Note
that, as detailed in that section, these error bars reflect not only the
precision of the series coefficients, but also the extreme sensitivity of the
Padé approximant in these coefficients in the presence of a defect.

The density of states (top panel) displays the usual Kondo effect features: a
thin Kondo peak at the Fermi level and lower and upper Hubbard bands centered
around $\omega = \pm U/2$. The agreement between the two methods is very
good, except at the tip of the Kondo peak. The Friedel sum
rule~\cite{Hewson1993} imposes that in the Kondo regime
$-\Im[G^R(\omega=0)] = 1 / \Gamma$. It is respected by QQMC (plain line), but not by FTPS
(dashed line) which lacks resolution at very low-frequencies due to its finite
time limit.

The middle and lower panels show respectively the real and imaginary parts of
the self-energy, where the agreement is also good. Deviations around
$\omega = \pm 3\Gamma$ are attributed, by elimination of other possibilities,
to the resummation error. We expect this to improve by increasing the number of
orders. At larger frequencies $|\omega| > 5\Gamma$, inaccuracies in the QQMC
integration are the cause of the disagreement, as can be seen by the larger
error bars. Although FTPS does not capture the low energy Green function
perfectly, it still captures the Fermi liquid features, as it is much more
precise in the low energy self-energy. Indeed, in our experience the
$\eta$-extrapolation works better in the self-energy than in
the spectral function. We speculate that this might be because the Kondo peak
is sharper than the low-frequency self-energy, and therefore more difficult to
extrapolate.

\begin{figure}[t]
   \centering
   \includegraphics[width=\columnwidth]{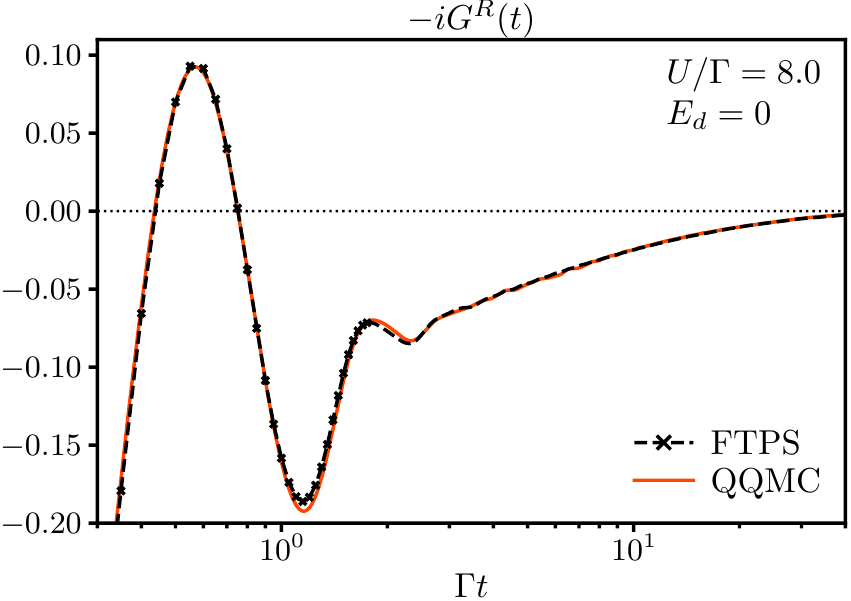}
   \caption{
      \label{fig:bench-time}
      Retarded Green function in time domain, in the $E_d = 0$ model at $U/\Gamma = 8$. FTPS (dashed
      line and symbols) yields valid result only for a limited
      time, after which finite bath size effects occur. Symbols are shown only at small
      times for readability. QQMC (plain line) works directly in the thermodynamic
      limit, but may be less accurate at short times (high frequencies). Due to
      particle-hole symmetry, $G^R(t)$ is
      pure imaginary.
   }
\end{figure}

\subsection{Symmetric model in time}

It is instructive to compare the Green function for the symmetric model in the
time domain, as shown in Fig.~\ref{fig:bench-time}. The Fourier transform of
FTPS data suffer from an additional error due to its finite time extent. Hence,
we show here the FTPS data before $\eta$-extrapolation (dashed
line and symbols, symbols are shown only at small times for readability). The
QQMC data (plain orange line) is the same as in Fig.~\ref{fig:bench-sym}
after Fourier transformation. Error bars have not been propagated through this
Fourier transform, as it would require knowledge of noise correlations between
frequencies, which has been lost during the Padé resummation.

The QQMC result shows good agreement with FTPS, concerning the large
oscillations and the long time decay rate, and out-range FTPS at long times
(note the logarithmic time scale). However, discrepancies can be seen in the
high frequency features. The difference in the two first oscillations is linked
to the mismatch in Fig.~\ref{fig:bench-sym} around $\omega \approx
\pm 3\Gamma$,
already discussed above. The small mismatch in the long time oscillations
($\Gamma t > 3$) is connected to frequencies near the bandwidth
$\omega \approx \pm D$, where QQMC has lower resolution when calculating
coefficients (see Fig.~\ref{fig:mcmc-vs-qqmc}).

\begin{figure}[b]
   \centering
   \includegraphics[width=\columnwidth]{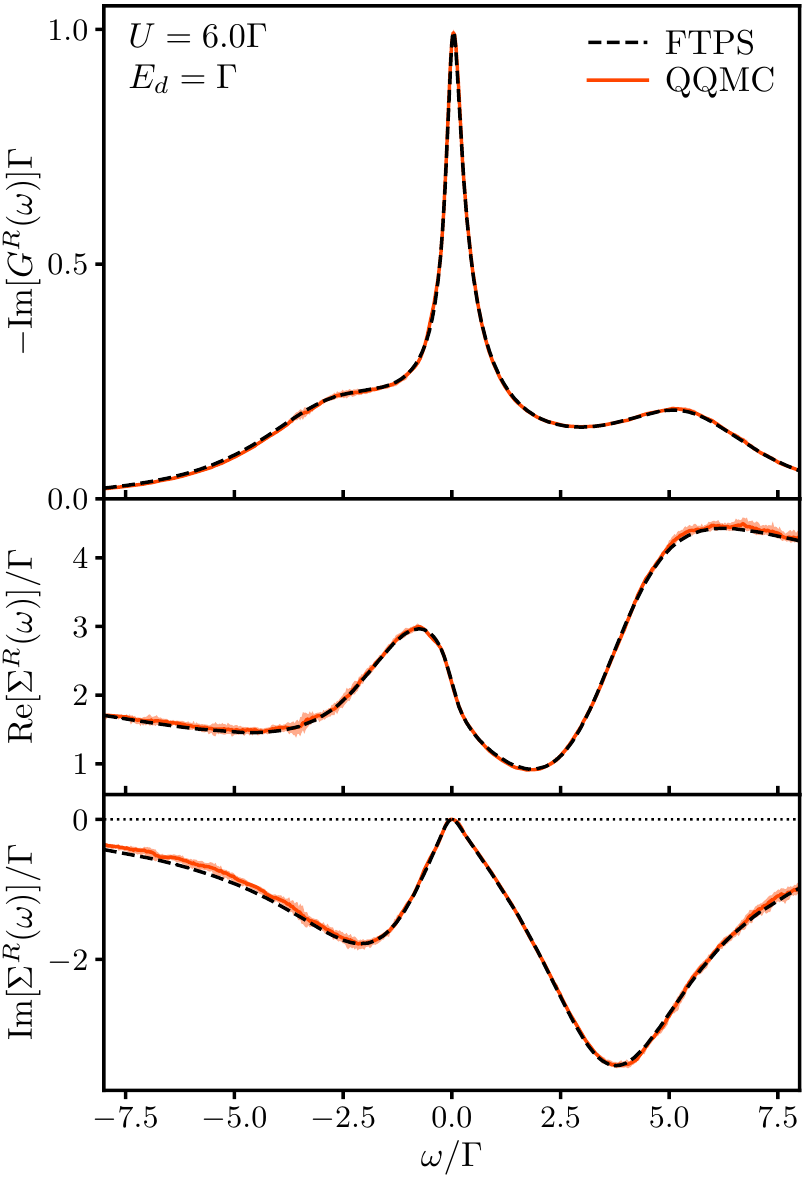}
   \caption{
      \label{fig:bench-asym}
      Comparison between QQMC result after resummation (plain line) and FTPS (dashed
      line), at $U=6\Gamma$  in the asymmetric case. Top panel: spectral function
      on the dot. Middle and bottom panels: real and imaginary parts of the
      associated self-energy. Shaded areas represent QQMC integration error
      estimation.
   }
\end{figure}

\subsection{Asymmetric model}

Finally, we consider the asymmetric model ($E_d = \Gamma$) at
$U = 6\Gamma$, for which the comparison in the frequency domain is
shown in Fig.~\ref{fig:bench-asym}. The resummation was done with the
$[6/6]$ Padé approximant at low frequencies
$|\omega| < 4\Gamma$, and $[4/4]$ at high frequencies. The
same progressive transition has been used as in the symmetric model.

In spite of the lower interaction, we still recognize the main features of the
Kondo regime in the density of states (upper panel): the Kondo peak at the
Fermi level and an upper and lower Hubbard bands at $\omega \approx E_d \pm U / 2$. As
the Kondo peak is broader than in Fig.~\ref{fig:bench-sym}, it is expected
to be better captured by FTPS than in the particule-hole symmetric case. The real and imaginary parts of the self-energy
(middle and lower panels) show an overall good agreement. As in the
particle-hole symmetric case, large frequencies are more noisy in the QQMC
result, due to larger relative errors when calculating coefficients.

\section{Conclusion}

Using the recently developed QQMC method based on low-discrepancy integration
of Ref.~\onlinecite{Macek2020}, we computed a full real-time Green
function perturbation series in an interacting quantum system. We compared two
different algorithms that adapt the kernel-based technique of
Ref.~\onlinecite{Bertrand_1903_kernel}, originally designed with a Markov chain Monte
Carlo integration. The first one computes the kernel at a given frequency as a
single integral, while the second computes its whole time dependence at once
using the same sampling of the integration space. We optimize the QQMC
integration by using a warping technique which introduces information on the
integrand in a problem-independent way.

For both methods, switching from traditional Monte Carlo to the novel QQMC
brought an important speedup in the calculation of the Green function
perturbation series. This is caused by an improved convergence, the error
scaling in the best cases as $1/N^{0.86}$, with
$N$ the number of samples. In practice, we typically gain
2-3 orders of magnitude in precision. More importantly, the switch to
quasi-Monte Carlo opened up the possibility to further improve the convergence
rate. Indeed, unlike with Monte Carlo, this rate depends on the smoothness of
the integrand, which could be improved by more advanced warpings.

The full kernel method turns out to be superior to the single frequency method,
strengthening the idea that a single sampling distribution can be used
efficiently to compute a continuum of correlators. Nevertheless, more advanced
warpings could change the ratio of performance in the future.

Applying this technique to a zero-temperature Anderson impurity model, and
after resummation of the Green function series using Padé approximants, we
compared the full kernel QQMC result to the non-perturbative FTPS technique.
This comparison brought an overall very good agreement between the two very
different methods. The observed discrepancies can be linked to limitations in
both algorithms: the low frequencies are better resolved by the kernel method
due to the long time limitation of FTPS, but the high frequencies are more
accurate in the FTPS results, probably due to biases introduced by the
resummation and integration noise.

From the FTPS viewpoint, this comparison showed that long time Green functions
results (up to $t=40/\Gamma$) using only time evolution, as well as
low frequency self-energies are reliable.

The QQMC technique is versatile and can easily be adapted to more complex
systems such as multi-band or multi-orbital impurity models, or lattice models,
although performance is still an open question. In addition, the integration algorithm
is highly automatic, thanks to the projection-based technique for building
tailor-made warpings.

Further developments can be made to improve the current algorithm for computing
the Green function perturbation series. First, high frequency noise could be
reduced by adapting QQMC to computing the $L$ kernel, as
defined in Ref.~\onlinecite{Bertrand_1903_kernel}. Finally, as with the calculation of a
single quantity, building warpings that capture more features of the integrand
would allow faster convergence, potentially allowing access to higher
perturbation orders.


\begin{acknowledgments}
   We would like to thank M.~Ferrero and F.~Šimkovic for useful discussions on
   Padé approximants. The algorithms in this paper were implemented using code
   based on the TRIQS library~\cite{TRIQS2015} and the QMC-generators
   library~\cite{Kuo_1606}. The Flatiron Institute is a division of the
   Simons Foundation. XW and MM acknowledge funding from the French-Japanese ANR
   QCONTROL, E.U.~FET UltraFastNano and FLAG-ERA Gransport.
\end{acknowledgments}

\appendix

\begin{figure*}[t]
   \centering
   \includegraphics[width=\linewidth]{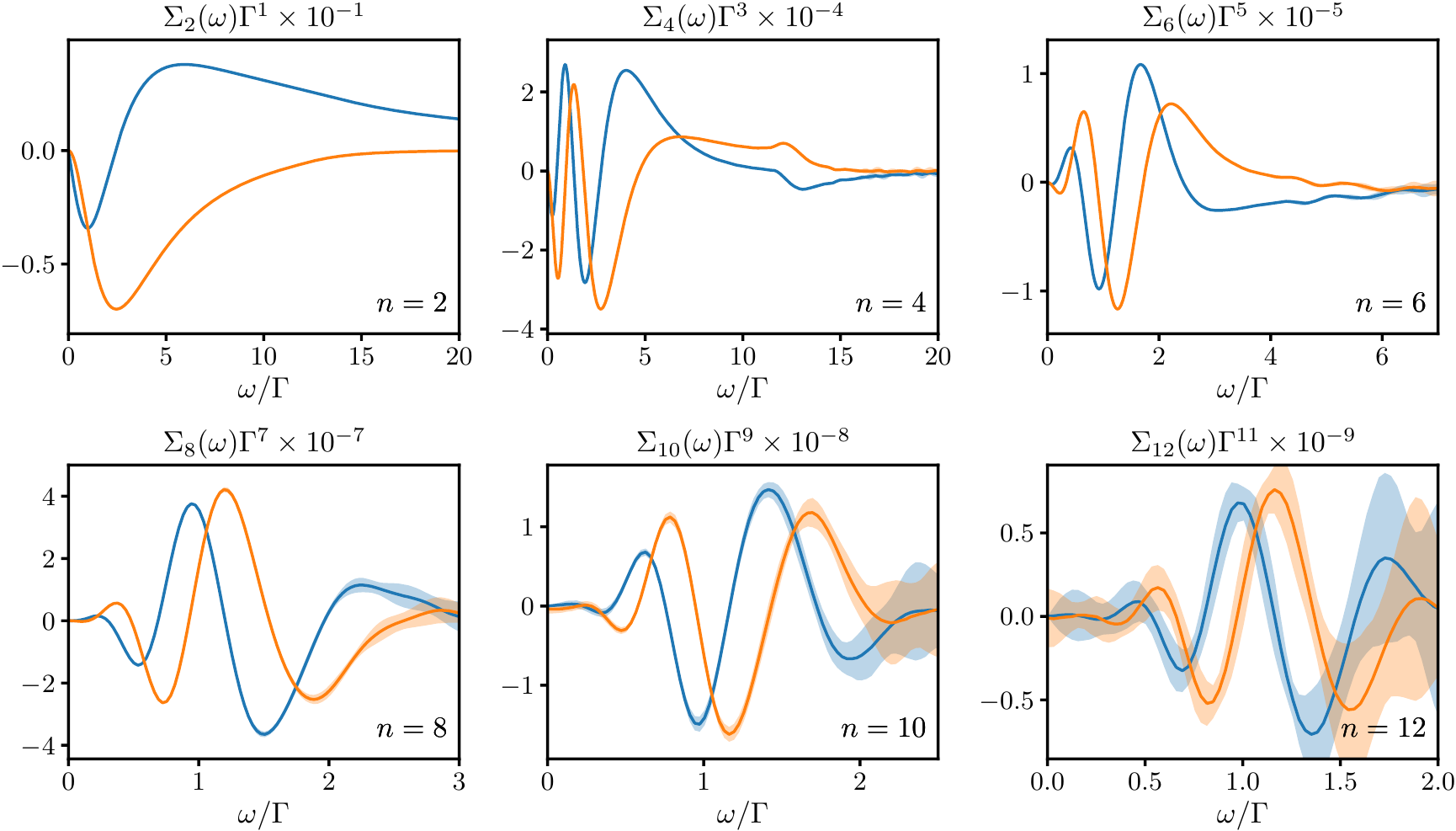}
   \caption{
      \label{fig:self-energy-all-sym}
      Perturbation series for $\Sigma(\omega)$ in the symmetric model
      ($E_d = 0$). Orders $n=2, 4, \ldots, 12$ are shown in reading
      order from the top left. The real part is in blue, the imaginary in orange. The estimated error is shown as a shaded area.
   }
\end{figure*}

\section{Equal-time Green function: a time splitting implementation}
\label{app:time-splitting}

As mentioned in Sec.~\ref{sec:definitions}, care needs to be taken when
evaluating the Wick determinant, since the Green functions
$g^{T}(t,t'), g^{\widetilde{T}}(t,t')$ have a discontinuity at $t = t'$. Here
we elaborate on how to correctly addressed this problem.

First, the Green functions $g(U_k, U_k)$ along the diagonal of the
determinant correspond to Wick contraction of Fermion operators within the
interacting Hamiltonian $H_I$. These Green functions are
replaced by $g^<(u_k, u_k) - i \alpha$. The choice $g^<$
reflects the correct operator ordering $\sim c^\dagger c$ for each spin
block of $H_I$; the $-i\alpha$ term accounts for
the quadratic shift.

Secondly, we encounter the situation where times $u = v$ in
Green functions, which correspond to Wick contractions between operators
between two Hamiltonians $H_I(u)H_I(v)$. The $u \to v$
limit is on the edge of the integration region in the time ordered perturbative
expansion and has measure zero. For numerical evaluation of the integral,
however, we want to include this boundary and define it such that the
$u \to v$ limit is smooth.

A simple way to implement this smooth limit is to impose a ``time-splitting''
procedure. Specifically, each time appearing in \eqnref{eq:integral_gf} is
associated with an additional splitting index\footnote{Note that $t$ and $t'$ don't need to be
distinguished, as $g^{ab}(t, t')$ appears only in disconnected diagrams at order
$>0$.}, which can
be appended to the combined index
\begin{align}
   X   & = (t, a) \rightarrow (t, a, 0),         \\
   X'  & = (t', b) \rightarrow (t', b, 0),       \\
   U_k & = (u_k, a_k) \rightarrow (u_k, a_k, k).
\end{align}
The definition of the non-interacting Green function $g$ is
adjusted so that for any $A = (u, a, s)$ and $B = (v, b, r)$,
\begin{equation}
   g(A, B) = \lim_{\epsilon \rightarrow 0} g^{ab}(u + s\epsilon,
   v + r\epsilon).
\end{equation}
In practice, the order of $A$ and $B$
on the Keldysh contour is determined: first, by the Keldysh indices
$a, b$; second, if $a=b$, by the times
$u,v$; third, if $a=b$ and
$u = v$, by the splittings $s,r$. Thus,
$A = B$ if and only if Keldysh indices, times and splittings
are equal. The Green function is then unambiguously
\begin{equation}
   g(A, B) =
   \begin{cases}
      g^<(u, v), & \mathrm{if}~A \le B, \\
      g^>(u, v), & \mathrm{otherwise}.
   \end{cases}
\end{equation}

\section{Perturbation series of the self-energy}
\label{app:series}

The perturbation series for the self-energy in the symmetric model
($E_d = 0$) is shown up to order 12 in
Fig.~\ref{fig:self-energy-all-sym}. The real part is in blue, the imaginary in
orange. The estimated error is shown as a shaded area. In this model, even
orders are zero due to the particle-hole symmetry.

The self-energy coefficients lose about one order of magnitude in amplitude
every 1--2 perturbation orders, providing good convergence properties. To
benefit from this, we need to compute coefficients with an amplitude as low as
$10^{-9}$ at order 12.

\section{Susceptibility of some Froissart doublets to noise}
\label{app:elim_defects}

\begin{figure}[t]
   \centering
   \includegraphics[width=\columnwidth]{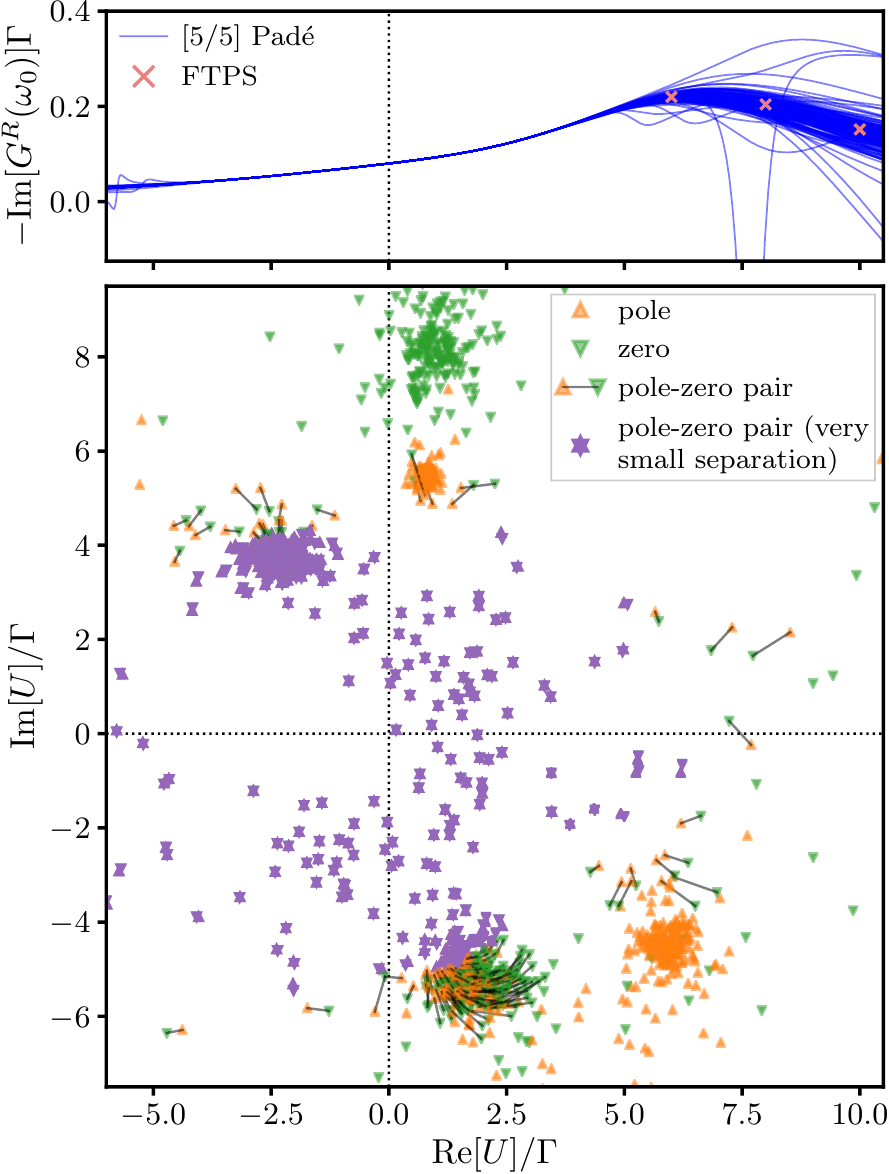}
   \caption{
      \label{fig:spurious-poles}
      Distribution of poles, zeros, and pole--zero pairs of the $[5/5]$ Padé approximant of a noisy
      series (bottom panel) and corresponding evaluation of the approximant on the
      real axis (top panel). The series is $G^R_n(\omega_0)$ for the asymmetric
      model ($E_d=\Gamma$) with addition of different realizations of a
      Gaussian noise compatible with QQMC error bars, with $\omega_0
      = -2.58\Gamma$. Poles are orange triangles pointing
      up, zeros are green triangles pointing down. Within each realization of
      noise, pole--zero pairs which are close (separation $\delta<\Gamma$) are shown
      with a black line joining them. Purple symbols shows pairs with a very small
      separation ($\delta < 0.2\Gamma$, overlapping symbols), a clear indication of a defect. These are spread out in a large region of the
      complex plane. Occurrences near the real axis are not rare, but affect much more the evaluation of the approximant (blue lines in top
      panel) in the range $\Re[U]/\Gamma > 4$ than $\Re[U]/\Gamma < 4$.
   }
\end{figure}

Froissart doublets or defects are known features of Padé approximants formed of
a pole and a zero in the complex plane, separated by a small distance
$\delta$. These defects strongly disrupt the expected behavior
of the approximant in their vicinity, but have a vanishing effect
$\sim \delta / L$ at long distance $L$. The location
of defects can be very sensitive to accuracy on the coefficients of the series.
This can be a problem, if such defects appear close to a region of interest. In
this appendix, we look at the influence of uncertainty on the coefficients
$G^R_n(\omega_0)$, on the location of defects and the evaluation of the
Padé approximant. For in-depth mathematical studies of the phenomenon on
simpler series, we refer to the literature \cite{Gilewicz1997,Gilewicz1999}.

We consider the asymmetric model ($E_d = \Gamma$) and the
$[5/5]$ Padé approximant for a given frequency
$\omega_0 = -2.58 \Gamma$. We generate a Gaussian noise in the series
$G^R_n(\omega_0)$ that is compatible with the QQMC error bars, ignoring
correlations between coefficients.

The location of the poles and zeros of the $[5/5]$ Padé
approximant using 200 realizations of this noise are displayed in
Fig.~\ref{fig:spurious-poles} (lower panel). Poles are shown as orange triangles  pointing up,
and zeros as green triangles pointing down. Poles and zeros that are suspected
to be part of a defect are displayed differently. If a pole and a zero of the
same Padé are close enough so that their symbols
overlap($\delta < 0.2\Gamma$) , they are drawn in purple. Otherwise, if
their separation is $\delta < \Gamma$, they are linked together by a
black line.

A $[5/5]$ Padé approximant has 5 poles and 5 zeros in the
complex plane. We can locate them in Fig.~\ref{fig:spurious-poles}. We notice
immediately two poles ($U/\Gamma \approx 6 - 4i$ and $U/\Gamma \approx 1 + 5.5i$) and
a zero ($U/\Gamma \approx 1.5 + 8i$) that are stable. In addition, two other stable
structures (at $U/\Gamma \approx 2 - 5i$ and $U/\Gamma \approx -2.5 + 4i$) are formed
of pole--zero pairs, and are probably defects. These are far from the real axis
so they do not affect strongly the evaluation of the Padé approximant there.
More interesting is the last pair that spans a large region of the complex
plane, in particular many occurrences (but not all) have a modulus
$<3/\Gamma$. This defect is problematic as it may appear very close
to the real axis. Finally, a last zero spreads mostly out of the shown area, at
very large moduli.

The top panel of Fig.~\ref{fig:spurious-poles} shows the spectral function (beam
of blue lines), evaluated on the real $U$ axis from the
above-mentioned Padé approximants, for each realization of noise. Notice the
stability of the evaluation for $|\Re[U]/\Gamma| < 4$, even though several
defects appear close to the real axis in this range. However, these defects
have an extremely small separation $\delta \sim 10^{-5}\Gamma$ compared to their
distance from the real axis $L \sim 10^{-1}\Gamma$. As $U$
increases, the beam of lines spread further more but stays consistent with FTPS
calculations (red crosses). However, a dozen among all 200 lines have dramatic
variations, inconsistent with the other lines or with FTPS. These are caused by
the defects observed close to the real axis in the range
$4 < \Re[U]/\Gamma < 8$. These defects have a larger separation
$\delta \sim \Gamma$, and therefore a higher probability to be within a few
$\delta$ from the real axis. As one can see, the defect causing
wild variations of the density of states has a strong probability to lie within
a distance $< 3\Gamma$ of the origin, where it is observed not to
affect evaluation on the real axis. We use this probability in
Sec.~\ref{sec:pade} to eliminate the effect of this defect.

The analytical structure of Padé approximants possesses features that are more
or less stable to perturbations in the Taylor coefficients. We observe that
some defects are extremely unstable and can vary wildly in location, whereas
some poles and zeros are stable.

\section{Empirical convergence of Padé approximants}
\label{app:pade-convergence}

\begin{figure}[t]
   \centering
   \vspace{0.2in}
   \includegraphics[width=\columnwidth]{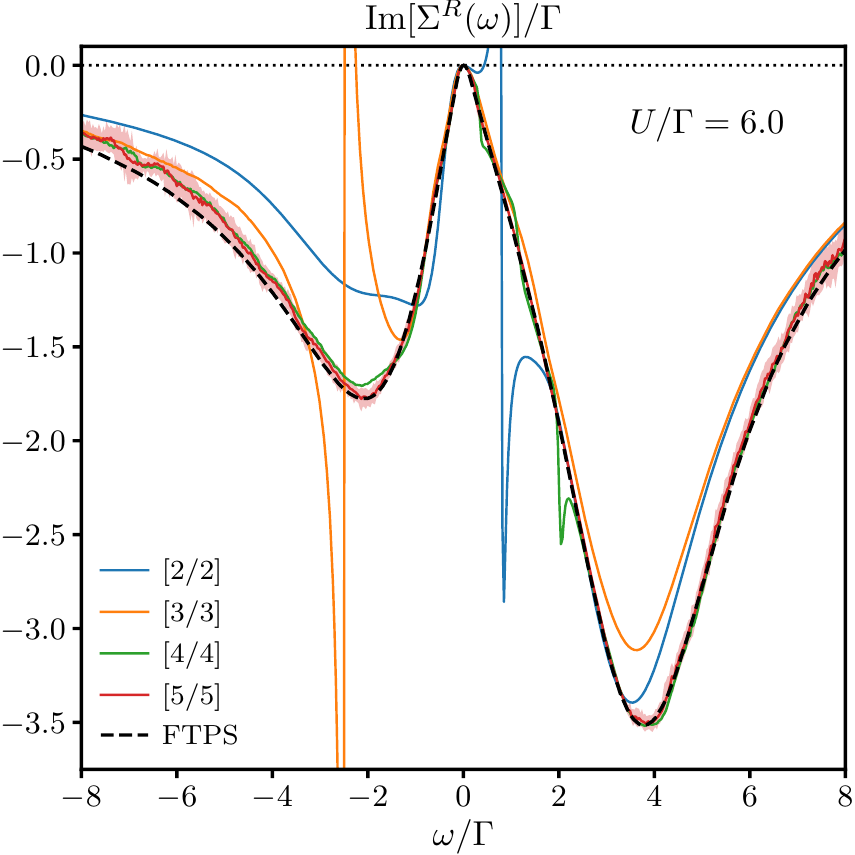}
   \caption{
      \label{fig:pade-converg}
      Convergence of the self-energy (imaginary part) using $[m/m]$
      Padé resummation of the Green function with increasing $m=2, \dots, 5$
      (plain lines). The model is the $E_d=\Gamma$ one at
      $U=6\Gamma$. Notwithstanding abrupt variations caused by defects in
      Padé approximants, the sequence converges toward a result consistent with the FTPS computation
      (dashed line). For clarity, only the error bar of $m=5$ is shown (pale red area).
      $m=2$ and $3$ errors are smaller than the
      line width,
      $m=4$ error is of the same magnitude as
      $m=5$.
   }
\end{figure}

The convergence of a sequence of $[l/m]$ Padé approximants with
increasing $l$ and $m$ is the subject
of intense mathematical research. No known result allows us to prove that the
approximants we consider in this work are part of a uniformly converging
sequence of functions.

However, we observe that several sequences of Padé approximants empirically
converge, ignoring spurious peaks caused by defects. Such a sequence is
represented in Fig.~\ref{fig:pade-converg}. The figure shows the imaginary
part of the self-energy at $U=6\Gamma$ obtained by resummation of
the series $G_n^R(\omega)$ using $[m/m]$ Padé
approximants (plain lines). We show $m=2, \ldots, 5$;
$m=6$ is not displayed for clarity, as it is difficult to
distinguish from $m=5$. The resummation follows the same
statistical treatment as described in Sec.~\ref{sec:pade}, to remove
the least stable defects. Some defects nevertheless survived, as can be seen
for instance around $\omega=-2.5\Gamma$ and $\omega=0.8\Gamma$, in the
$[3/3]$ and $[2/2]$ Padé approximants
respectively (strong variations in the self-energy are correlated to strong
variations in the Green function).  Ignoring these extreme variations, the
successive approximants seem to converge toward a result that is consistent
with the FTPS calculation (dashed line).

It is interesting to note that the $[5/5]$ (as well as
$[6/6]$) approximant seems free of defects, whereas lower order
approximants are not. It is possible that the larger uncertainty in the
evaluation of large order $G^R_n(\omega)$ make defects more susceptible
to our statistical treatment, and as a result easier to erase.

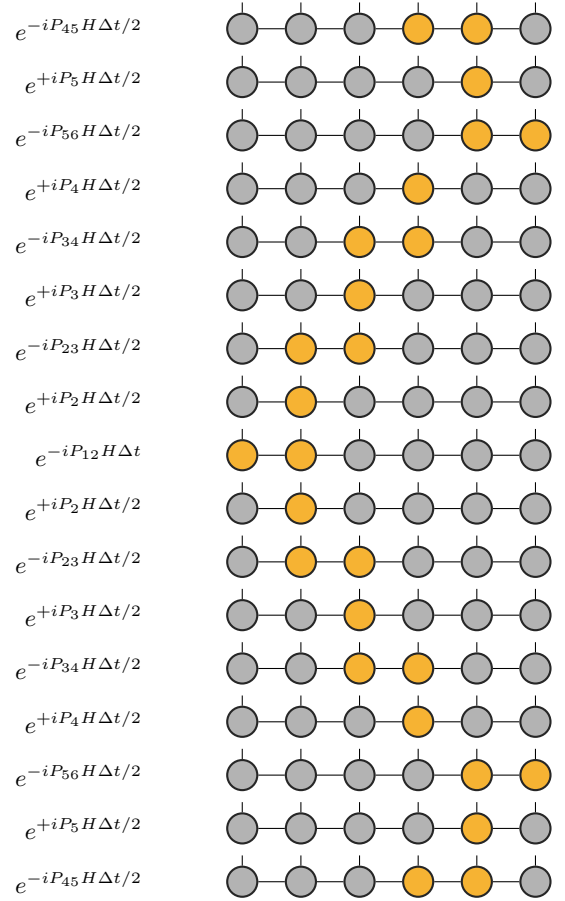
\begin{figure}[b]
   \centering
   \begin{tikzpicture}
    
   \definecolor{DanBlue}{RGB}{66,133,244}
   \definecolor{DanRed}{RGB}{219,68,55}
   \definecolor{DanGreen}{RGB}{15,157,88}
   \definecolor{DanYellow}{RGB}{244,160,0}

    \tikzstyle{old}=[	circle, thick, minimum size=0.4cm,
    				draw=black!85, fill=black!30, inner sep=0pt,  text width=4mm]

     \tikzstyle{new}=[	circle, thick, minimum size=0.4cm,
    				draw=black!85, fill=black!30, inner sep=0pt, text width=4mm]
    
   \tikzstyle{act}=[	circle, thick, minimum size=0.4cm,
   				draw=black!85, fill=DanYellow!80, inner sep=0pt, text width=4mm]
    
   \tikzstyle{Imp}=[	circle, thick, minimum size=0.4cm, draw=black!85,
				fill=DanGreen!80, inner sep=0pt, text width=4mm]
				
   \tikzstyle{Bath}=[	circle, thick, minimum size=0.4cm, draw=black!85,
				fill=black!30, inner sep=0pt, text width=4mm]
        
    \matrix (m) [below=1.5cm of m2, matrix of nodes, column sep=10, row sep = 8] {
       |[old]| & |[old]| & |[old]| & |[act]| & |[act]|& |[old]|\\
       |[old]| & |[old]| & |[old]| & |[new]| & |[act]|& |[old]|\\
       |[old]| & |[old]| & |[old]| & |[new]| & |[act]|& |[act]|\\
       |[old]| & |[old]| & |[old]| & |[act]| & |[new]|& |[new]|\\
       |[old]| & |[old]| & |[act]| & |[act]| & |[new]|& |[new]|\\
       |[old]| & |[old]| & |[act]| & |[new]| & |[new]|& |[new]|\\
       |[old]| & |[act]| & |[act]| & |[new]| & |[new]|& |[new]|\\
       |[old]| & |[act]| & |[new]| & |[new]| & |[new]|& |[new]|\\
       |[act]| & |[act]| & |[new]| & |[new]| & |[new]|& |[new]|\\
       |[old]| & |[act]| & |[new]| & |[new]| & |[new]|& |[new]|\\
       |[old]| & |[act]| & |[act]| & |[new]| & |[new]|& |[new]|\\
       |[old]| & |[old]| & |[act]| & |[new]| & |[new]|& |[new]|\\
       |[old]| & |[old]| & |[act]| & |[act]| & |[new]|& |[new]|\\
        |[old]| & |[old]| & |[old]| & |[act]| & |[new]|& |[new]|\\
       |[old]| & |[old]| & |[old]| & |[new]| & |[act]|& |[act]|\\
        |[old]| & |[old]| & |[old]| & |[new]| & |[act]|& |[old]|\\
        |[old]| & |[old]| & |[old]| & |[act]| & |[act]|& |[old]|\\
    };
    
   \foreach \z in {1,2,3,4,5,6,7,8,9,10,11,12,13,14,15,16,17} {
    	\foreach \x/\y in {1/2,2/3,3/4,4/5,5/6} \draw (m-\z-\x) -- (m-\z-\y);
	\foreach \x in {1,2,3,4,5,6} \draw (m-\z-\x) -- ($(m-\z-\x)+(0,0.35)$); 
   }
   
   \node[left=of m-1-1](a){~$e^{-iP_{45}H\Delta t/2}$};
   \node[left=of m-2-1](b){$e^{+iP_{5}H\Delta t/2}$};
   \node[left=of m-3-1](c){~$e^{-iP_{56}H\Delta t/2}$};
   \node[left=of m-4-1](d){$e^{+iP_{4}H\Delta t/2}$};
   \node[left=of m-5-1](e){~$e^{-iP_{34}H\Delta t/2}$};
   \node[left=of m-6-1](f){$e^{+iP_{3}H\Delta t/2}$};
   \node[left=of m-7-1](g){~$e^{-iP_{23}H\Delta t/2}$};
   \node[left=of m-8-1](h){$e^{+iP_{2}H\Delta t/2}$};
   \node[left=of m-9-1](i){~$e^{-iP_{12}H\Delta t}$};

   \node[left=of m-10-1](h){$e^{+iP_{2}H\Delta t/2}$};
    \node[left=of m-11-1](g){~$e^{-iP_{23}H\Delta t/2}$};
    \node[left=of m-12-1](f){$e^{+iP_{3}H\Delta t/2}$};
    \node[left=of m-13-1](e){~$e^{-iP_{34}H\Delta t/2}$};
    \node[left=of m-14-1](d){$e^{+iP_{4}H\Delta t/2}$};
    \node[left=of m-15-1](c){~$e^{-iP_{56}H\Delta t/2}$};
    \node[left=of m-16-1](b){$e^{+iP_{5}H\Delta t/2}$};
    \node[left=of m-17-1](a){~$e^{-iP_{45}H\Delta t/2}$};

\end{tikzpicture}
   \caption{
      \label{fig:sweep}
      Depiction of the TDVP sweeping order the time step to time evolve from time
      $t$ to $t+\Delta t$. This picture is the
      equivalent of Eq.~\ref{eq:TDVP_usualorder} and shows the terms that are being
      integrated on the left. The steps shown start in the middle of the MPS (on the
      impurity site) and move outwards the spin-up bath (see also
      Fig.~\ref{fig:MPS}). Yellow dots are the sites that are updated in
      each step. As usual, two-site updates are in forward direction (negative exponent) while single-site
      updates are in backward direction (positive exponent). Note that this is quite
      different from the usual TDVP given by Eq.~\ref{eq:TDVP_usualorder} which would
      start with the middle term ($P_{12}$) move
      upwards, but would do steps (1--3) in a different order. See also
      Ref.~\onlinecite{BauernfeindTDVPTTN} for more details.
   }
\end{figure}

\section{TDVP Time Evolution}
\label{app:TDVP}

The main idea behind the Time Dependent Variational Principle (TDVP) is to find
the best possible representation of time evolved states represented as MPS. To
do so it solves a modified Schr\"odinger equation in which the right-hand side
is changed: $H| \psi  \rangle \to P H | \psi  \rangle$. The projection operator
$P$ projects onto the so-called tangent space of the
current MPS $| \psi \rangle$ and keeps the time integration within the
manifold spanned by $| \psi \rangle$~\cite{HaegemanUnifyingTevo}. In the
two-site variant of TDVP, $P$ is given
by~\cite{HaegemanUnifyingTevo}:
\begin{equation}
   P = \sum_{i=1}^{N-1} P_{i,i+1}  -
   \sum_{i=2}^{N-1} P_i,
\end{equation}
with so-called two-site projection operators $P_{i,i+1}$ and
single-site projectors $P_{i}$. The exact form of these
operators is of no relevance here and can be found in
Ref.~\cite{HaegemanUnifyingTevo}. Their sole purpose is to solve the Schr\"odinger
equation only in the subspace spanned by the MPS. The two-site projectors
result in the usual forward time propagation, while the single site projectors
stem from the gauge degree of freedom of the MPS and make sure that entries are
not time evolved twice. This is achieved via a backwards time evolution with
opposite sign to the two-site projectors, see also Eq.~\ref{eq:TDVP_formalSolution}.
Importantly this means that the formal solution of the modified Schr\"odinger
equation is given by:
\begin{equation}
   | \psi(t+\Delta t) \rangle = e^{-i \left( \sum_{i=1}^{N-1} P_{i,i+1}  - \sum_{i=2}^{N-1} P_i \right ) H \Delta t} |
   \psi(t) \rangle.
   \label{eq:TDVP_formalSolution}
\end{equation}
Every single term (e.g.  $\expHdt{P_{i,i+1}}$ or $e^{i P_{i+1} H \Delta t}$)
can easily be integrated nearly exactly using Krylov matrix exponentiation but
the whole sum is far too complicated to deal with at once. Hence a second order
trotter decomposition is used to split the whole exponential into manageable
parts usually starting with the two-site term containing
$P_{1,2}$ then $P_2$, next
$P_{2,3}$ etc.:
\begin{widetext}
   \begin{align}
      \expHdt{\left( \sum_{i=1}^{N-1} P_{i,i+1}  - \sum_{i=2}^{N-1} P_i \right )} \approx
      \expHdthalf{ P_{1,2}} \cdot \expHdt{ \left( \sum_{i=2}^{N-1} P_{i,i+1}  - \sum_{i=2}^{N-1} P_i \right )} \cdot \expHdthalf{ P_{1,2}}  \nonumber                                                                    \\
      \approx \expHdthalf{ P_{1,2}} \cdot e^{+i P_{2} H \Delta t}  \cdot \expHdt{ \left( \sum_{i=2}^{N-1} P_{i,i+1} - \sum_{i=3}^{N-1} P_i \right )} \cdot e^{+i P_{2} H \Delta t} \cdot \expHdthalf{ P_{1,2}} \nonumber \\
      \approx \cdots \approx\left( \prod_{i=1}^{N-2}  e^{-i P_{i,i+1}H \Delta t/2} \cdot e^{+i P_{i+1}H \Delta t/2} \right ) e^{-i P_{N-1,N}H \Delta t}\left( \prod_{i=N-2}^{1} e^{-i P_{i,i+1}H \Delta t/2} \cdot e^{+i P_{i+1}H \Delta t/2} \right ) \label{eq:TDVP_usualorder}.
   \end{align}
\end{widetext}
Note that the indices of the products above are defined such that the rightmost
as well as the leftmost term is the one that contains $P_{1,2}$.
Applying each operator in the order indicated by Eq.~\ref{eq:TDVP_usualorder}
results in the usual TDVP integration scheme as proposed in
Ref.~\onlinecite{HaegemanUnifyingTevo}. It is worth noting that starting the
trotterization with the $P_{1,2}$-term is a convenient choice
since it results in an algorithm very similar to DMRG but there is nothing
preventing one from starting with any other term. \\ ~ \\ In fact here, we use
a different integration order as shown in Fig.~\ref{fig:sweep} which is
the pictorial equivalent of Eq.~\ref{eq:TDVP_usualorder}. We start at one of the
center sites which is the impurity where the creation/annihilation operator is
applied (sites $4$ and $5$ in
Fig.~\ref{fig:sweep}), sweeping right, jumping back to the center and
sweeping left. Again, since we use a second order breakup, all steps have to be
applied twice in an order resulting from repeated second order trotter breakups
similar to Eq.~\ref{eq:TDVP_usualorder}.

We choose this different integration order because a direct application of
Eq.~\ref{eq:TDVP_usualorder} would lead to large but unnecessary errors,
especially for large values of $\Delta t$. If one were to use
exclusively Eq.~\ref{eq:TDVP_usualorder}, the first few time steps would have to
work with an inadequate basis, because after the application of the creation
(annihilation) operator the remaining basis consists only of states in which
the impurity is completely full (empty). It is known that TDVP is very
susceptible to a too small number of basis states (bond
dimension)~\cite{Mingru_TDVPKrylovExp} and therefore this inadequate basis leads to
large errors in the first few time steps. The scheme shown in
Fig.~\ref{fig:sweep} does not have this problem, since it can produce
the missing basis states in the very first step (the
$P_{4,5}$-term in Fig.~\ref{fig:sweep}). We stress that
the new scheme is important only for the first few time steps.
Eq.\ref{eq:TDVP_usualorder} is perfectly adequate, although not better than the
scheme of Fig.11, for larger times. Another reason for using this integration
order is that it is easier to generalize to multi-orbital problems which is the
main purpose of the FTPS tensor network.

\bibliography{biblio-qqmc-kernel.bib}

\end{document}